\documentclass[prb,onecolumn,showpacs,floatfix,epsf,epsfig]{revtex4}
\usepackage{epsf}
\usepackage{epsfig}
\usepackage{epstopdf}
\usepackage{graphicx}
\begin{document}
\title{Probing the possibility of coexistence of martensite transition and half-metallicity in Ni and Co-based full Heusler Alloys : An ab initio Calculation}
\author{Tufan Roy$^{1}$, Dhanshree Pandey$^{1}$ and Aparna Chakrabarti$^{1,2}$}
\affiliation{$^{1}$HBNI, Raja Ramanna Centre for Advanced Technology, Indore - 452013, India} 
\affiliation{$^{2}$ISUD, Raja Ramanna Centre for Advanced Technology, Indore - 452013, India} 

\begin{abstract}
 Using first-principles calculations based on density functional 
theory, we have studied the mechanical, electronic, and magnetic 
properties of Heusler alloys, namely, Ni$_{2}BC$ and Co$_{2}BC$ 
($B$ = Sc, Ti, V, Cr and Mn as well as Y, Zr, Nb, Mo and Tc; 
$C$ = Ga and Sn). On the basis of electronic structure (density 
of states) and mechanical properties (tetragonal shear constant), 
as well as magnetic interactions (Heisenberg exchange coupling 
parameters), we probe the properties of these materials in detail. We calculate the formation energy of these alloys in the (face-centered) cubic austenite structure to probe the stability of all these materials. From the energetic point of view, we have studied the
possibility of the electronically stable
 alloys having a tetragonal phase lower in 
energy compared to the respective cubic phase. A large number of the
magnetic alloys is found to have the cubic phase as their ground state. On the other hand, 
 for another class of alloys, the tetragonal phase has been found to have lower
 energy compared to the cubic phase. Further, 
we find that the values of tetragonal shear constant show a consistent 
trend : a high positive value for materials not prone to
tetragonal transition and low or negative for others. 
In the literature, materials, which have been seen to undergo 
the martensite transition, are found to be metallic in nature.
We probe here if there is any Heusler alloy which has a tendency to 
undergo a tetragonal transition and at the same time possesses 
a high spin polarization at the Fermi level. 
From our study, it is found
that out of the four materials, which exhibit 
a martensite phase as their ground state, three of these, namely, 
Ni$_{2}$MnGa, Ni$_{2}$MoGa and Co$_{2}$NbSn have a metallic
nature; on the contrary, Co$_{2}$MoGa 
 exhibits a high spin polarization.
\end{abstract}

 %Other ferromagnetic metals and alloys

\maketitle

\section{Introduction}   

Heusler alloys are intermetallic compounds with interesting 
fundamental properties and possible 
practical applications. Heusler alloys 
are typically known to be of two types: full-Heusler alloys (FHA) 
and half-Heusler alloys (HHA). The full-Heusler alloys, which are 
having a formula $A_{2}BC$ (with a stoichiometry
 2:1:1, e.g. Ni$_{2}$MnGa), 
commonly exhibit a $L2_{1}$ structure. This has four interpenetrating 
face-centered (fcc) sub-lattices, for each 
of the atoms $A$(Ni), $A$(Ni), $B$(Mn) and $C$(Ga). On the other hand,
 the� half-Heusler alloys with a formula $ABC$ (with a 
stoichiometry 1:1:1, 
e.g. NiMnSb) typically assume a $C1_{b}$ structure where one of the 
four fcc sub-lattices remains unoccupied.  Since the discovery of the 
prototype FHA, Ni$_{2}$MnGa\cite{PMB49PJW}, 
various studies have been carried out which show that this alloy, 
having long-range ferromagnetic interaction, possesses various interesting 
physical properties. For example, Ni$_{2}$MnGa exhibits magnetic 
field induced strain (MFIS) and magnetic shape memory alloy (MSMA) 
property\cite{Ullako96,Murray00,Sozinov02}, magnetoresistance effect 
(MRE)\cite{APL86CB} as well as magnetocaloric effect 
(MCE)\cite{APL76FXH}. 

 The Heusler alloys are interesting from both
the points of view of possible technological application as well as 
fundamental science. Hence, these have been enjoying the attention of 
the researchers - theoreticians and experimentalists alike. Further, the basic drawbacks of the prototype Heusler alloy, Ni$_{2}$MnGa, in terms of technological application are its brittleness and the low martensite transition temperature.  
Therefore, following the discovery of and studies on Ni$_{2}$MnGa, various FHAs have been synthesized, characterized and 
studied in the last two decades, as is observed in the literature.  
Many new FHA materials, till date, have been predicted from 
ab initio calculations as well.
It has been observed that the face-centered-cubic phase is 
the high-temperature or the so-called austenite 
phase of these materials. Some of these FHAs has been seen to 
undergo a tetragonal distortion at a lower temperature. This 
first-order displacive transition, generally known as         
 martensite transition, where the volumes of the unit cell of both
the austenite and martensite phases are close to each other, is 
typically connected to the SMA property exhibited by these alloys. 
Among the FHAs, in terms of the electronic structure, there are
various categories. While some of these alloys prefer to be 
metallic, some are found to be semiconducting, and some are 
having a large spin-polarization at the Fermi 
level.\cite{PMB49PJW,electronic-str,PRL50RAdG,PT54WEP} 

If the FHAs are magnetic in nature, their properties can change
when a magnetic field is applied which can be of interest in terms of 
potential technological application. Hence, specially, magnetic shape 
memory alloys are gaining increasing interest. Therefore, detailed 
studies of magnetic configurations, properties and interactions are 
of particular importance. In literature, various magnetic ground 
state configurations are observed in case of full-Heusler alloys.
While some alloys are even non-magnetic, many of these exhibit a long-range
 ferromagnetic ordering and are expected to show MSMA property. In many of 
these MSMAs, there is presence of a delocalized-like common 
d-band formed by the d-electrons of the $A$ and $B$ atoms, which are 
both typically first-row transition metal atoms.\cite{PRB28JK}
 Additionally, there is also an 
indirect RKKY-type exchange mechanism\cite{RKKY}, primarily 
mediated by the electrons of the $C$ atoms, 
which plays an important role in defining the magnetic properties
of these 
materials.\cite{PRB28JK,PRB77ES,PMB49PJW,PRB72ATZ,JALCOM632TR} 
Further, it has also been observed that some of these alloys including Mn$_{2}$NiGa even 
show long-range ferrimagnetism and also anti-ferromagnetism.\cite{APL87liu,EPL80bar,PRB87KRP,APL98IG,JMMM401TR} 

Hence, it is clear from the literature, that in terms of different
physical, including, structural (mechanical), electronic and 
magnetic properties, the full-Heusler alloys show a rich variety. 
Further, as has been mentioned above, it has been of particular 
interest that out of all the full-Heusler 
alloys, only a few undergo the martensite transition. These alloys 
are prone to a cubic to tetragonal distortion when temperature is 
lowered and generally exhibit the technologically important SMA 
property.  These FHAs in general are found to be metallic in 
nature. On the other hand, it has been observed that there is another
group of full-Heusler alloys which are half-metallic-like in nature, 
with a much reduced density of states (DOS) at the Fermi level 
in case of one of the 
spin channels. These materials generally do not 
show the tendency of undergoing a tetragonal distortion  
 and also showing the SMA property. However, 
an application in the field of spintronics is a possibility
for these materials.
From both the points of view of fundamental understanding as well
as technological application, 
it can be interesting to probe the similarities and 
differences in magnetic, bulk mechanical, and electronic properties 
of these two categories of materials. It will also be 
interesting to see if there is any FHA which has a tendency to 
undergo a tetragonal transition and at the same time possesses 
a high spin polarization at the Fermi level.

Keeping this in mind, 
in the present paper, we probe the magnetic, bulk mechanical, 
 and electronic properties of a series of Ni and Co-based
full Heusler alloys using density functional theory
 (DFT) based ab initio calculations. 
The choice of these two systems (Ni and Co-based FHAs) is due to the 
following facts. First and foremost, it has been seen that typically,
a large amount of work on the FHAs are on Ni and Co-based compounds.
It is also seen in the literature that
while most of the Ni-based FHAs show MSMA property, many of the
Co-based FHAs exhibit large spin-polarization at the Fermi level.
It has also been pointed out in the literature, that while the 
magnetic interactions are somewhat different in the Ni and Co-based 
FHAs, the total energy variation for an austenite to martensite phase
transition is similar.\cite{AEM14MS} 
Hence, a comparative study may
be interesting and also important for detailed understanding of the
properties of these alloys. The primary interest is to study the 
possibility of the tetragonal transition versus a high spin polarization
at the Fermi level. Further, we look for ferromagnetic 
materials so that realization of MSMA property is possible.
In what follows, first, we give a brief account of the methods we used 
and then we present the results and discussion. In the end, the 
results of this work are summarized and conclusions are drawn.

\section{Method}  
The full-Heusler alloys, as for example, Ni$_{2}$MnGa, commonly 
assume an ordered $A_{2}BC$ structure, where typically $A$, $B$ are 
elements with d-electrons and $C$ are elements with s,p electrons.
In the cubic high-temperature austenite phase, Ni$_2$MnGa has 
a $L2_{1}$ structure that consists of four interpenetrating
 face-centered-cubic (fcc) sub-lattices with origin at fractional 
positions,
(0.25,\,0.25,\,0.25),
 (0.75,\,0.75,\,0.75)
(0.5,\,0.5,\,0.5), and
(0.0,\,0.0,\,0.0).
In $L2_1$ structure of Ni$_2$MnGa, the Ni atoms occupy the first and second sub-lattices. On the contrary, Mn and Ga occupy the third and fourth sub-lattices,
respectively. In this paper, we have carried out calculations on Ni 
and Co-based systems. So Ni and Co are taken as $A$ atom and 
$C$ = Ga as well as Sn. 
As for the $B$ atom, we have taken into consideration and consequently 
tested the electronic stability of the first five atoms of the first
as well as second rows of the transition metal atoms. Therefore, Sc, 
Ti, V, Cr and Mn as well as Y, Zr, Nb, Mo and Tc are considered as 
the $B$ atom.

The equilibrium lattice constants of all these alloys are obtained by full 
geometry optimization using Vienna Ab Initio Simulation Package 
(VASP)\cite{VASP} which has been used in combination
with the projector augmented wave (PAW) method\cite{PAW} and the 
generalized gradient approximation (GGA) over the local density 
approximation (LDA) for the exchange-correlation functional.\cite{PBE} 
GGA is used because it accounts for the density gradients, and hence, 
for most of the Heusler alloy systems, it has been found that it 
yields results which are in better agreement with experimental data   
 compared to the results of LDA. We have used an energy cutoff of 
500\,eV for the planewaves. The final energies have 
been calculated with a $k$ mesh of 15$\times$15$\times$15 for the 
cubic case and a similar number for the tetragonal case. The energy and 
the force tolerance for our calculations 
were 10 $\mu$eV and 10 meV/\AA, respectively. The formation energies ($E_{form}$), as calculated\cite{VASP} by the 
equation below,
has been critically analysed to establish the electronic stability 
of the alloys.
\begin{equation}
E_{form} = E_{tot} - \Sigma_{i} c_{i}E_{i}
\end{equation}
where $i$ denotes different types of atoms present in the unit cell 
of the material system and $E_{i}$ are the standard state energies of
the corresponding atoms $i$.\cite{VASP}
The optimized geometries of the systems studied are compared with the 
results obtained in the literature, wherever results are available, 
and these match well with earlier data as discussed in the section on 
Results and Discussion.

The response of a material to an applied stress is associated with
the elastic constants of the material.  Both stress ($\sigma$) and 
strain ($\epsilon$) in a material have three tensile as well as three
shear components. Therefore, the linear elastic constants form
a 6$\times$6 symmetric matrix. We have $\sigma_{i}$ = $C_{ij}$ 
$\epsilon_{j}$ for small stresses, $\sigma$, and strains, $\epsilon$.
Calculations of the mechanical properties of the 
materials involve the variation of total energy of 
the system induced by the strain.\cite{VASP}
Elastic constants of all the materials are evaluated from the second 
derivative of the energy with respect to the strain tensor.
The number of
$k$-points and the energy cut-off have been increased from the values
used in SCF
calculations till the convergence of the mechanical properties of
each individual material has
been achieved. Mesh of $k$-points has been taken as
15$\times$15$\times$15 and energy cut-off of
500 eV as per the convergence requirement.

It is well known that, all-electron calculations are more reliable for the prediction of magnetic properties particularly for the systems containing first row transition elements. Hence, to calculate and understand in detail the magnetic as well as electronic properties, we have carried out relativistic
spin-polarized all-electron calculations of all the systems, geometries being optimized by VASP.\cite{VASP} These calculations have been performed using full
potential linearized augmented planewave (FPLAPW) 
program\cite{Wien2k} with the generalized gradient approximation 
(GGA) for the exchange-correlation functional.\cite{PBE}
For obtaining the electronic properties, the Brillouin zone (BZ) 
integration has been carried out using the tetrahedron method 
with Bl\"ochl corrections.\cite{Wien2k}
An energy cut-off for the plane wave expansion of about 17 Ry is used
($R_{MT}$$K_{max}$ = 9.5).  The cut-off for charge density is 
$G_{max}$= 14. The number of $k$ points for the self-consistent 
field (SCF) cycles in the reducible (irreducible) BZ is about 8000
 (256) for the cubic phase and about 8000 (635) for the tetragonal 
phase. The convergence criterion for the total energy $E_{tot}$ is 
about 0.1 mRy per atom. The charge convergence is set to 0.0001. 

To gain further insight into the magnetic interactions of some of the 
magnetic materials, we calculate and discuss their Heisenberg exchange 
coupling parameters. The geometries optimized by VASP have been used for these calculations. We use the Spin-polarized-relativistic 
Korringa-Kohn-Rostoker method (SPR-KKR) to calculate the Heisenberg 
exchange coupling parameters, J$_{ij}$, within a real-space approach, 
which is proposed by Liechtenstein et al\cite{JMMM67AIL} 
and implemented in the SPR-KKR programme package.\cite{sprkkr} 
The mesh of k points for the SCF cycles has been taken as 
21$\times$21$\times$21 in 
the BZ. The angular momentum expansion for each atom is
taken such that lmax=3. In addition, in terms of the Heisenberg 
exchange coupling parameters we derive the Curie temperature 
(T$_{C}$) following the literature\cite{JPCM23MM}.

\section{Results and Discussion}  

\subsection{Geometry Optimization and Electronic Stability} 

{\bf Lattice parameter and Atomic number $Z$ of $B$ atom} - For 
the cubic phase, the 
$L2_{1}$ structure has been assumed for all the structures studied
here, namely, Ni$_{2}BC$  and Co$_{2}BC$  ($B$ = Sc, Ti, V, Cr 
and Mn as well as Y, Zr, Nb, Mo and Tc; $C$ = Ga and Sn). The geometry
 has been optimized to obtain the converged lattice parameter.
Figure 1 and Figure 2 show the variation of this lattice parameter 
as a function of $Z$ of $B$ elements of $A_{2}BC$ alloy ($A$ = Ni, Co; 
$C$ = Ga, Sn). The $B$ atoms correspond to the period IV of
the periodic table (first row transition metal atoms; Sc etc) and 
the period V (second row transition metal
atoms; Y etc). Therefore, in Figures 1 and 2, we have mentioned 
the period numbers IV and V in the legends.
For the Ni$_{2}BC$ materials, it is observed, as the atomic number of 
$B$ elements increases, with a saturating trend towards higher $Z$, 
lattice parameters of the materials decrease for a fixed row of the 
periodic table (Figure 1). This may happen due to 
the increasing electro-negativity and decreasing atomic radius of 
atoms across the column. Further, for these materials we observe 
(in the left panel of Figure 1) a sudden increase
in the lattice parameter value for $Z$ of $B$ = 24 (i.e. Cr atom). 
It is to be noted that out of all the materials studied here, a 
deviation from the ferromagnetic nature is expected for Ni$_{2}$CrSn 
as well as Ni$_{2}$CrGa which was reported earlier.\cite{JMMM401TR}
These two materials have lower energy for a long-range 
{\it intra-sublattice} anti-ferromagnetic (AFM) ordering compared 
to the ferromagnetic (FM) one. This difference in the long-range 
magnetic interaction between the FHAs with $B$ = Cr and the 
other $B$ atoms indicates that the Cr-based materials are possibly 
of a different class compared to the rest of the FHAs, considered here. 
This may be the reason behind the deviation from the observed trend. 
It is to be further noted that for the AFM ordering the lattice 
parameter (shown by a black square) is even larger compared to
the corresponding FM phase. However, the long-range magnetic
interaction in all the materials with $A$ = Co is ferromagnetic.
 For these Co-based materials the 
lattice parameters show a smooth decrease as we increase the $Z$ of 
the $B$ atom (Figure 2). As discussed above for Ni-based systems, 
this may, again, be due to the increase of 
electronegativity and decrease of atomic radius across the column 
which may have led to the shrinkage of the electron cloud around the 
$B$ atom, and consequently, of the whole unit cell. It has been 
observed that there is an increase in the lattice parameter values 
when we go from period IV to period V, for both Ni and Co-based 
materials. Increase of atomic radius is observed across the row, 
going from period IV to V, and the above-mentioned 
trend may be because of that. 

{\bf Formation energy and Atomic number $Z$ of $B$ atom} - Figure 3 
as well as Figure 4 suggest that the formation energy is negative for 
all of the materials with $C$ atom = Ga, except Ni$_{2}$TcGa which 
is having marginally positive formation energy. It is to be noted
that Co$_{2}$TcGa has marginally negative formation energy and hence
in reality may not be stable electronically. It is well-known that 
negative 
formation energy signifies stability; more negativity indicates 
more stability of the material. A few of the materials, $A_{2}BC$, 
where $C$ = Sn, like Ni$_{2}$MoSn, Ni$_{2}$TcSn, Co$_{2}$CrSn, 
Co$_{2}$MoSn and Co$_{2}$TcSn, are having positive formation energy.
These calculated values and the prediction that these 
particular materials are electronically unstable, matches with
the results wherever available in the literature.\cite{MGThesis} 
The AFM phases for Ni$_{2}$CrGa and Ni$_{2}$CrSn have very close but
slightly smaller $E_{form}$ compared to the respective FM case. 
It is observed from both the figures that, overall, there is a trend 
 of electronic stability decreasing as $Z$ of $B$ atom increases. 
However, for $B$ = Mn, the stability has increased  
compared to the previous $B$ element. This suggests an interesting
 preferance for the Mn atom in the $B$ position for the Heusler 
alloys  
with $L2_{1}$ structure in both the cases when $A$ atom is Ni or Co. 
Similar is the case with $B$ = Zr.
It is seen that, for Y, i.e. the first atom of the second row of 
the transition metal atoms (period V) at the $B$ position, 
 the formation energy is somewhat unfavorable compared 
to the next case of $B$ = Zr. The origin of this has been found to
be electronic in nature - analysis of the density of states 
for $B$ = Zr indicates a lowering of binding energy in this
material compared to the $B$ = Y case. It is observed that the 
contribution from the majority spin density of states of the Co
atom plays an important role. For our further studies, we 
concentrate only on the materials which, from our calculations,
 are found to be electronically stable.

{\bf Electronic Stability of the Tetragonal phase} - We calculate
the difference between the energy of the cubic ($E_{C}$) and 
tetragonal ($E_{T}$) phases 
of all the electronically stable materials. Figure 5 shows this 
energy difference, $\Delta E$ = $E_{T}$ - $E_{C}$ (in units of meV 
per atom), of some typical materials 
as a function of the ratio of lattice constants $c$ and $a$. 
From our calculations, we find that only a few materials, among all 
the electronically stable and magnetic FHAs we study here,
exhibits the tetragonal phase as a lower energy state. Among all the 
Ni-based alloys, we find that Ni$_{2}$MoGa and 
Ni$_{2}$CrSn possess a lower energy for the tetragonal phase 
over the cubic phase similar to 
 Ni$_{2}$MnGa and Ni$_{2}$CrGa, both of which have a tetragonal ground 
state, as has already been shown in the literature. 
From Figure 5, we also observe that 
Ni$_{2}$VGa and Ni$_{2}$VSn have very flat energy curves with 
no clear minimum in the $\Delta E$ versus $c$/$a$ curve. 
Further, we observe that though in case of Ni$_{2}$MnSn, there is 
a cubic ground state, there is also a very subtle signature of a 
tetragonal phase which is evident from the clear asymmetric nature of
the $\Delta E$ versus $c$/$a$ curve for this material (Figure 5).
It is observed that 
 Ni$_{2}$MoGa exhibits a non-magnetic state as ground state. 
Further, Ni$_{2}$CrGa and 
Ni$_{2}$CrSn are likely to possess an {\it intra-sublattice} AFM 
phase as a ground state. It is to be noted that, in this paper, our 
concentration will be only on the alloys which will have FM phase in its 
ground state. So, for Ni-based alloys, further on, we will discuss only 
about Ni$_{2}$MnGa and Ni$_{2}$MnSn. 

Out of all the Co-based alloys studied here, we observe that only two 
alloys are likely to show energetically lower martensite phase over
the cubic austenite phase. Out of these two, while Co$_{2}$NbSn 
is known in the literature\cite{JPSJ58SF}, Co$_{2}$MoGa is not 
reported till date. 
From Figure 5, it is clearly seen that for Co$_{2}$MoGa a significant 
energy difference exists between the cubic and tetragonal phases. 
This indicates that a martensite phase transition is possible in 
this material. It is expected that the martensite transition 
temperature for Co$_{2}$MoGa will be higher than Ni$_{2}$MnGa
since the energy difference $\Delta E$ between the austenite and
martensite phase of the former is evidently much larger than 
that of the latter. This expectation is because it is argued in the 
literature, that, relative values of $\Delta E$ can be indicative 
of the relative values of martensite transition temperatures of 
 two alloys.\cite{AEM14MS,PRB78SRB} 
A cubic ground state is observed for many Co-based alloys, including
the two well-known Co-based half-metallic-like materials 
Co$_{2}$MnSn\cite{JALCOM645MB} and Co$_{2}$MnGa\cite{PRB71YK}. 
It is interesting to note
that there are three more alloys which show a state of cubic symmetry 
having a lowest energy while the tetragonal phases of these 
materials are energetically very close (within 25 meV) 
to the respective austenite 
phases. The $\Delta E$ versus $c$/$a$ plots of these materials, 
namely, Co$_{2}$VGa, Co$_{2}$CrGa and Co$_{2}$TcGa are included 
in Figure 5 which clearly depict this energetic aspect.

\begin{table*}
Table~1. Martensite transition temperature of the four ferromagnetic materials\footnote{Comparison with the experimental or theoretical data, wherever results are available \\
$^{b}$Ref.\onlinecite{JMMM401TR}
$^{c}$Ref.\onlinecite{JPCM-25-025502-2013}
$^{d}$Ref.\onlinecite{APL86CB}
$^{e}$Ref.\onlinecite{PRB-71-174428-2005}}
\begin{tabular}{|c|c|c|c|c|}
\hline Material&$(c/a)_{eq}$&$\Delta E$(meV/atom) &$T_{M}$(K)&$|\Delta{V}|$/V( \%)\\
\hline Ni$_{2}$MnGa&1.22&6.05&70.18&0.30\\
&$1.22^{b}$, $1.22^{c}$&$6.18^{b}$&$70.51^{b}$, $210^{d}$&\\
\hline Ni$_{2}$MoGa&1.27&11.08&128.58&0.36\\
\hline Co$_{2}$MoGa&1.37&19.58&227.13&1.96\\
\hline Co$_{2}$NbSn&1.11&10.63&123.33&0.08\\
&&&$233^{e}$&\\
\hline
\end{tabular} 
\end{table*}

In Table 1 we report the tetragonal transition temperature ($T_{M}$) of 
those materials which are expected to exhibit tetragonal transition. 
These are calculated based on $\Delta E$ using the conversion factor 
1 meV = 11.6 K. As discussed above, these $T_{M}$ values are not 
to be considered as the absolute values of the transition 
temperature. These values are listed here to only give a trend of 
the relative transition temperatures for different materials as has been
done in the literature earlier.\cite{AEM14MS,PRB78SRB} 
Out of the four ferromagnetic materials (magnetic aspect of the 
materials is discussed in detail in the next subsection) 
showing the possibility of a tetragonal transition, 
we find that Ni$_{2}$MnGa
has the lowest $T_{M}$ value; on the other hand, Co$_{2}$MoGa
is expected to have the highest transition temperature. The 
optimized $c$/$a$ values for all the four materials are given in  
Table 1. Co$_{2}$MoGa is found to have the highest value
of about 1.4. Furthermore, volume conservation between the cubic and the 
tetragonal phases (Table 1) as well as an energetically lower tetragonal 
phase (Figure 5) which are observed here are generally indicative of the 
martensite transition. Therefore, from the present study, among the 
materials studied here, two Ni-based and two Co-based FHAs are likely to 
exhibit MSMA property. Out of which one Ni and
one Co-based alloys are well-known MSMA materials, namely Ni$_{2}$MnGa
and Co$_{2}$NbSn. On the other hand, Co$_{2}$MoGa and Ni$_{2}$MoGa
are two new materials which are also likely to exhibit martensite 
transition.

\subsection{Magnetic Properties}

\begin{table*}[ht]
Table~2. Magnetic properties of austenite phase of Co-based materials; 
$Z_{t}$ is the total number of valence electrons\footnote{Comparison with experiments or previous calculations, wherever data are available \\
 $^{b}$Ref.\onlinecite{JMMM-163-313-1996};
  $^{c}$Ref.\onlinecite{JPCS-35-1-1974};
   $^{d}$Ref.\onlinecite{JALCOM-317-406-2001};
    $^{e}$Ref.\onlinecite{JMMM-30-374-1983};
     $^{f}$Ref.\onlinecite{PRB-76-024414-2007};
      $^{g}$Ref.\onlinecite{PRB-72-214412-2005}
      $^{h}$Ref.\onlinecite{ACTA-57-2702-2009}
       $^{i}$Ref.\onlinecite{JMMM-25-90-1981}
        $^{j}$Ref.\onlinecite{JPD40HCK}}
\begin{tabular}{|c|c|c|c|c|c|c|c|}
\hline  Material    &$\mu_t$&$Z_{t}$-24&$\mu_A$&$\mu_B$&$\mu_C$&$T_C$(K)&P(\%)  \\ 
\hline Co$_{2}$ScGa & 0.00&0 &0.00&0.00&0.00& -&- \\ 
&0.25$^{b}$&&&&&&\\
\hline Co$_{2}$TiGa &1.00 &1 &0.62 &$-$0.14&$-$0.01&161&97\\ 
&0.75$\pm 0.03$$^{c}$,0.82$^{d}$ && 0.40$^{b}, $0.40$\pm 0.1$$^{c}$ &&& 130$\pm3$$^{c}$,128$^{d}$&\\
\hline Co$_{2}$VGa  &2.00 &2 &0.95&0.18&$-$0.01&356&100 \\ 
&1.92$^{e}$, 2.00$^{f}$&&0.91$^{f}$&&&352$^{e}$, 368$^{f}$& \\
\hline Co$_{2}$CrGa &3.03 &3 &0.77  &1.59&$-$0.05&418&92\\
&3.011$^{g}$, 3.01$^{g}$&&0.90$^{g}$&1.28$^{g}$&$-$0.07$^{g}$&419$^{g}$, 495$^{g}$&\\ 
\hline Co$_{2}$MnGa &4.10 &4 &0.77  &2.73&$-$0.07&586&68\\ 
&4.04$^{h}$&&&&&700$^{h}$&\\
\hline Co$_{2}$YGa  &0.00 &0 &0.00&0.00&0.00& -&- \\ 
\hline Co$_{2}$ZrGa &1.00 &1 &0.61&$-$0.11&$-$0.01&166&94  \\ 
\hline Co$_{2}$NbGa &2.00 &2 &1.04 &$-$0.01&0.00&397&100 \\
&1.39$^{i}$, 2.00$^{j}$ &&&&&&\\ 
\hline Co$_{2}$MoGa &2.93 &3 &1.22&0.51 &$-$0.01 &180&86 \\ 
\hline Co$_{2}$TcGa &3.95 &4 &1.37&1.26&$-0.04$&711&71 \\ 
\hline Co$_{2}$ScSn &1.05 &1 &0.67&$-$0.14&$-$0.02& 207&80 \\ 
\hline Co$_{2}$TiSn &2.00 &2 &1.07&$-$0.06&0.00& 409&100 \\ 
&1.96$^{e}$&&&&&371$^{e}$&\\
\hline Co$_{2}$VSn  &3.00 &3 &1.08&0.89&$-$0.02 & 291&100 \\ 
&1.21$^{e}$, 1.80$^{f}$ &&&&& 95$^{e}$, 103$^{f}$ &\\
\hline Co$_{2}$MnSn &5.03 &5 &0.98&3.23&$-$0.06& 897&76\\ 
&5.02$^{f}$&&0.885$^{f}$&3.25$^{f}$&&899$^{f}$&\\
\hline Co$_{2}$YSn  &1.05 &1 &0.67&$-$0.10&$-$0.02&162&79 \\ 
\hline Co$_{2}$ZrSn &2.00 &2 &1.10  &$-$0.09&0.01&449&100 \\ 
&1.46$^{e}$&&&&&448$^{e}$&\\
\hline Co$_{2}$NbSn &1.98 &3 &0.97 &0.07&0.01&37& -66 \\ 
&0.69$^{e}$&&&&&105$^{e}$&\\
\hline 
\end{tabular} 
\end{table*}
 
{\bf Total and Partial Magnetic Moments} - 
After analysing in detail the electronic stability of the 
austenite phase as well as the relative energetics of the
martensite phase of the materials, 
 we now discuss in detail the magnetic properties of the cubic 
phase of the Ni and Co-based materials, only
 which are ferromagnetic in nature. The materials which are electronically 
unstable are not discussed further as well.
As observed earlier\cite{JMMM401TR} for Ni$_{2}$CrGa, it is seen
that Ni$_{2}$CrSn too is expected to show {\it intra-sublattice} 
anti-ferromagnetism; though in both cases, partial moment on Cr atom 
is seen to be significant. Further, a first-principles calculation 
by Sasioglu et al on a series of materials, like Pd$_{2}$MnZ, 
Cu$_{2}$MnZ, show that the magnetic moment is mainly confined on the 
Mn sublattice for these alloys which contain Mn as the $B$ atom, and 
a very small moment is induced on the Pd or Cu atom.\cite{PRB77ES}  
Similarly, in the Ni-based materials studied here, 
 the magnetism in this class of materials is expected to arise 
primarily due to the $B$ element. This is because,
 by itself, Ni carries a very small 
magnetic moment and, Ga as well as Sn are having almost zero moment.
Often the moment of the $A$ atom is seen to be strongly influenced by 
the $B$ atom as is observed in the literature\cite{PRB77ES,PRB88AC} as 
well as for the Ni-based alloys studied here, except the $B$ = Mn 
materials, namely, Ni$_{2}$MnGa and Ni$_{2}$MnSn. 
 Ni$_{2}$VSn, which has a very flat minimum in the $\Delta E$
versus $c$/$a$ curve, has a very small moment as well. We find that, 
out of all the Ni-based materials, only two Ni$_{2}BC$ alloys, namely,
 Ni$_{2}$MnGa and Ni$_{2}$MnSn exhibit the FM nature 
and they have very similar total moments. While the total moment is 
4.10 $\mu_{B}$ in the former, it is 4.09 $\mu_{B}$ in the latter. 
Each of Ga and Sn has negligible moments in both cases. As for Ni
atom, the moment on this atom is larger in the former alloy (0.36)
compared to the latter material (0.24). Further, it is noted that
Mn atom has a much larger moment in case of Ni$_{2}$MnSn (3.64)
compared to the case of Ni$_{2}$MnGa (3.41). Though the common
$B$ atom has a difference of moment of 0.23 $\mu_{B}$, due to the
reasonably lower moment of the Ni atom in case of Ni$_{2}$MnSn, 
the total moments of these two materials are found to be very 
close to each other.\cite{PRB70ES,JPCM11AA} These findings can 
be supported by the observed larger lattice parameter of Ni$_{2}$MnSn 
compared to that of Ni$_{2}$MnGa. 
Due to the larger lattice parameter in the former, the
delocalization of the $3d$ electrons of Mn atom is expected to 
decrease,
 leading to a larger and more atomic-like partial moment on the same.  
The larger lattice constant 
in Ni$_{2}$MnSn leads to the decrease in the overlap between the 
Mn and Ni atoms, as is 
evident from the relative DOS too, as discussed later. This
may be the likely reason as to why the moment of the Ni atom 
decreases in case of Ni$_{2}$MnSn in comparison to 
Ni$_{2}$MnGa, as discussed in the literature.\cite{JPCM11AA}

Table 2 gives the total and partial moments for the electronically 
stable Co-based materials. The values available from the literature 
are also listed in the table for a few materials, wherever 
available and we note that the 
matching is very good with the existing calculated data. With
the experimental results the matching is reasonably good. It is seen
that each of Ga and Sn has negligible moments in all cases.
We observe that, as opposed to the materials with 
$A$ = Ni atom, those with $A$ = Co atom have significant contribution 
from the $A$ atom to the total moment. However, when the $B$ atom 
is non-magnetic, in a few cases, the moment on the Co atom is zero or
much less compared to its bulk moment. As the moment on the $B$ atom 
increases, the moment of the Co atom gets larger but always largely          
underestimated compared to the value of its bulk moment (about 
1.7$\mu_{B}$). This is expected due to the delocalized-like common d-band
between the $A$ and $B$ atoms.\cite{PRB28JK} 
For period IV, when $B$ = Cr and Mn, there is a slight 
decrease in the partial moment of the Co atom, 
but not in the total moment value of the respective systems. For all 
the Co-based alloys, the moments are very close to an integer 
value and this is generally the signature of a half-metallic material.
Further, from Table 2, we note that these FHAs follow the 
Slater-Pauling rule as is seen earlier 
in case of many Co-based FHAs.\cite{JPD40HCK,PRB66IG}
As a consequence of this rule, we get an almost linear variation of the 
magnetic moment with the atomic number of $B$ elements for all the 
Co-based materials. It is seen that 
there is a deviation from the Slater-Pauling rule only for Co$_{2}$NbSn
which has been observed and explained in the literature.\cite{JPD40HCK} 
Due to the electronic structure, all the Co-based compounds are seen 
to exhibit (Table 2) a high spin 
polarization at the Fermi level ($E_{F}$) in comparison to the Ni-based
compounds. From our calculations, Ni$_{2}$MnGa and Ni$_{2}$MnSn have 
spin-polarizations $\sim$28 and $\sim$21\%, respectively. 

{\bf Heisenberg Exchange Coupling Parameters} - To gain insight into the magnetic interactions in detail we calculate and  
present the results of our calculation of the Heisenberg exchange
coupling parameters, $J_{ij}$, ($i$ and $j$ being pairs ($A$, $A$) and 
($A$, $B$)) for the alloys which are likely to exhibit tetragonal distortion 
and consequently martensite transition. We also show the
same parameters for some other related alloys for the purpose of 
comparison. Most of the materials chosen for presentation have relatively
large moment on the $B$ atom so that the ($A$, $B$) exchange interactions 
are always significant. In the left panels of Figures 6, 7 and 8, 
 the $J_{ij}$ parameters for different compounds are plotted. 
The right panels give the interaction parameters, $J_{ij}^{bare}$, 
which are $J_{ij}$ parameters, divided by
the product of the moments of the $i$ and $j$ atoms.

Figure 6 gives these parameters for Ni$_{2}$MnGa and Ni$_{2}$MnSn 
which exhibit 
the effect of the change of the $C$ atom, and consequently the lattice 
parameter. Figure 7 contains the values for Co$_{2}$MnGa and 
Co$_{2}$MoGa to understand what is the role of the $B$ = Mn over 
the Mo atom. Similarly, we plot the exchange parameters for alloys 
Co$_{2}$MnSn and Co$_{2}$NbSn in Figure 8. 
The results match well with the literature
wherever the data are available.\cite{PRB70ES,PRB88AC} 
It is seen that there is a RKKY\cite{RKKY} type of interaction
for the ($A$, $A$) and ($B$, $B$) pairs.\cite{PRB28JK,PRB77ES}
 The oscillatory behavior of the $J_{ij}$ 
parameters as a function of distance between the atoms $i$ and
$j$ (normalized with respect to the lattice constant), 
is a well-known signature of the same. Further, it is seen that 
between the $A$ and the $B$ atoms there is a signature of 
a significant direct interaction whenever $B$ has a strong moment. 
From Figure 6, we
observe that the direct interaction between Ni-$B$ atom is stronger
in case of Ni$_{2}$MnGa compared to the case of Ni$_{2}$MnSn. This is 
due to the increased lattice constant and hence weak coupling 
in case of the latter. It is also found that the direct 
interaction is somehwat stronger than the RKKY interaction in case
of both the materials, though it is quite 
clear that, as expected and observed in the 
literature\cite{PRB28JK,PRB77ES}, both these 
types of magnetic interactions are important in these materials. 

The magnetic interactions in the Co-based materials shown 
here exhibit a somewhat similar trend. It is found that for 
the materials favoring tetragonal transition, 
Co$_{2}$MoGa (Figure 7) and Co$_{2}$NbSn (Figure 8), 
the direct $A$-$B$ magnetic interaction is small and to some extent
comparable to the indirect RKKY-type interaction between $A$ atoms. In the 
former material, as is evident from Table 2, the moment on Mo atom 
is somewhat larger compared to that on the Nb atom in case of the 
latter material.
Consequently, the strength of the direct interaction in the former alloy, 
between $B$ atom (Mo) and $A$ atom (Co) is found to be slightly more.
From Figures 7 and 8, we observe that for the 
materials Co$_{2}$MnGa and Co$_{2}$MnSn, having high magnetic 
moments (4.10 and 5.03 $\mu_{B}$, respectively), which are
 of the order of the moments 
possessed by the two Ni-based materials discussed above, the 
direct $A$-$B$ interaction is much stronger compared to the RKKY-type 
indirect interactions ($A$-$A$ or $B$-$B$). 
A decrease of partial moment of the Mn atom is observed in case of 
Co$_{2}$MnGa over Ni$_{2}$MnGa (2.73 in Co$_{2}$MnGa versus 3.41 $\mu_{B}$ 
in Ni$_{2}$MnGa). But due to larger moment on the $A$ = Co over Ni atom, 
the direct interaction strength between Mn and
the respective $A$ atoms, is much larger in the former material, 
as seen from top left panels of Figures 6 and 7.
Hence, the delocalized common 
$3d$ band between Mn and Co atom in case of Co$_{2}$MnGa is expected 
to be more effective compared to the case of Ni$_{2}$MnGa. 
We observe from Figures 6,7 and 8 that, relatively, 
 the RKKY-type indirect interactions are slightly stronger for the 
 $C$ = Sn over Ga atom. It is to be noted here that Sn atom has one valence 
electron more than Ga.

It is known that the magnetic interactions in the Heusler alloy materials
having a large moment on $B$ atom, comprise of a large contribution from 
the $A$-$B$ direct interaction. At the same time, contribution of 
the the $A$-$A$ and $B$-$B$ 
indirect RKKY type of interaction is important as well. 
The materials, which have high moment on the $B$ atom, 
 typically exhibit a large $A$-$B$ direct interaction when compared to 
the strength of RKKY-type interaction. 
However, when the $J_{ij}^{bare}$ parameters are analyzed, 
for the majority of the materials, it is seen 
that, both the direct $A$-$B$ interaction 
and RKKY-type interaction between 
 $i$ and $j$ ($i$ and $j$ being $A$ and $B$) atoms, 
are somewhat similar in strength. 
This observation reiterates the fact that the magnetic exchange 
interactions are not only the function of $i$-$j$ distances (as is 
evident from Figures 6 to 8), but also of
the individual magnetic moments of the $i$ and $j$ atoms which
becomes clear when we present the $J_{ij}^{bare}$ plots.

Based on the $J_{ij}$ parameters of the six materials discussed
above, the Curie temperatures of the materials within a mean-field 
approximation\cite{JPCM23MM} have been calculated to probe further
into the possible MSMA property. For Ni$_{2}$MnGa
and Ni$_{2}$MnSn, the Curie Temperature ($T_{C}$) values 
are 410 and 365 K, respectively. These values as well as the $J_{ij}$
parameters match quite well with
both experimental and calculated values reported in the
literature.\cite{PRB70ES,PRB88AC} We note that 
the experimental values are generally underestimated compared to 
the theoretical values. This is because of the
mean-field-approximation.  
For some of the Co-based materials $T_{C}$ values are presented in 
Table 2. We observe that the calculated values match very well with 
the previously reported data, wherever these results are available. 
It is interesting to note that due to the small values of $J_{ij}$
for Co$_{2}$NbSn the value of the Curie temperature for this material 
is very low and this is consistent with the experimentally 
observed room-temperature paramagnetism in this material. Similarly,
due to weak RKKY-type interaction of pair ($A$, $A$) with a somewhat 
comparable $A$-$B$ direct interaction, the $T_{C}$ value for 
another Co-based material, which has been predicted from our present work, 
 namely, Co$_{2}$MoGa, is expected to be below room temperature as well.

\subsection{Bulk Mechanical Properties}

\begin{table*}[ht]
Table~3. Bulk mechanical properties of austenite phase of Ni-based materials\footnote{Comparison with experiments or previous calculations, wherever data are available \\
$^{b}$Ref.\onlinecite{PRB54JW}
$^{c}$Ref.\onlinecite{JALCOM511SA}
$^{d}$Ref.\onlinecite{JPCM11AA}}
\begin{tabular}{|c|c|c|c|c|c|c|c|c|}
\hline Material&C$_{11}$(GPa )&C$_{12}$(GPa )&C$_{44}$(GPa)&C$\prime$(GPa)&B(GPa)&G$_{V}(GPa)$&G$_{R}(GPa)$&G$_{V}$/B \\
\hline Ni$_{2}$VGa&193.20&183.32&109.36&4.94& 186.62&67.59&11.56&0.36\\
\hline Ni$_{2}$MnGa&165.41&159.45&113.67&2.98&161.44&69.39&7.16&0.43\\
&152.0$^{b}$&143$^{b}$&103$^{b}$&4.5$^{b}$&146$^{b}$&63.6$^{b}$&&\\
\hline Ni$_{2}$MoGa&197.36&206.36&103.40&-4.5&203.36&60.24&-12.04&0.30\\
\hline Ni$_{2}$MnSn&161.02&137.46& 92.56&11.78&145.31&60.25&24.73&0.41\\
&158.1$^{c}$ &128.5$^{c}$&81.3$^{c}$, 87$^{d}$&14.8$^{c}$, 8$^{d}$&138.4$^{c}$, 140$^{d}$&&&\\
\hline
\end{tabular} 
\end{table*}

\begin{table*}[ht]
Table~4. Bulk mechanical properties of austenite phase of Co-based materials\footnote{ Comparison with experiments or previous calculations, wherever data are available \\
$^{b}$Ref.\onlinecite{PRB82TK} 
$^{c}$Ref.\onlinecite{JPCM11AA}}
\begin{tabular}{|c|c|c|c|c|c|c|c|c|}
\hline Material&C$_{11}$(GPa )&C$_{12}$(GPa )&C$_{44}$(GPa)&C$\prime$(GPa)&B(GPa)&G$_{V}(GPa)$&G$_{R}(GPa)$&G$_{V}$/B \\
\hline Co$_{2}$VGa&266.52 &162.12&126.83&52.20&196.92&96.98&80.68&0.49\\
&&&&&198$^{b}$&&&\\
\hline Co$_{2}$CrGa&233.02&182.82&136.77&25.1&199.56&92.10&49.20&0.46\\
\hline Co$_{2}$MnGa&254.87&165.27&142.69&44.80& 195.14&103.53&76.14&0.53\\
&&&&&199$^{c}$&&&\\
\hline Co$_{2}$MoGa&180.92&163.60&114.10&8.66&169.38&71.92&19.44&0.42\\
\hline Co$_{2}$TcGa&249.54&186.10&123.87&31.72&207.25&87.01&57.29&0.42\\
\hline Co$_{2}$MnSn&234.63&136.75&119.05&48.94&169.38&91.01&75.68&0.54\\
\hline Co$_{2}$NbSn&164.95&184.99&80.80&-10.02&178.31&44.47&-30.77&0.25\\
\hline
\end{tabular} 
\end{table*}

After presenting the results of magnetic properties of the austenite phase of the materials, 
now, we discuss about the bulk mechanical properties of the same  phase since from technological application point of view, these properties are important. 
For demonstrating the differences, we concentrate on eight typical FM materials and 
present the detailed results of the same. 
Three of these FM materials are likely to undergo a tetragonal 
distortion at low temperature and these materials are
Ni$_{2}$MnGa, Co$_{2}$MoGa, Co$_{2}$NbSn. The other group of 
alloys consists of materials having a cubic ground state. Among these 
alloys, the following compounds have been considered - Co$_{2}$VGa, 
Co$_{2}$CrGa, Co$_{2}$MnGa, Co$_{2}$MnSn and Ni$_{2}$MnSn, 
which are expected to have a cubic symmetry at the lowest temperature. 
Tables 3 and 4 contain the bulk mechanical properties of these 
above-mentioned Ni and Co-based alloys, respectively. 
Values of few other materials are also listed in Tables 3 and 4, 
for comparison. It is observed that the overall agreement with the values
from the literature is reasonably good.

 There are three independent elastic 
constants for a cubic structure. These are C$_{11}$, C$_{12}$ and 
C$_{44}$. These three elastic constants can be found by 
calculating energies for three different types of strain on the 
unit cell of the system under equilibrium. 
From these three 
linearly independent energy versus strain data, we can find 
out C$_{11}$, C$_{12}$ and C$_{44}$. 
The applied strains have the form as ($\delta$, $\delta$, $\delta$, 
0, 0, 0), (0, 0, 0, $\delta$, $\delta$, $\delta$) 
and ($\delta$, $\delta$, (1+$\delta$)$^{-2}$-1, 0, 0, 0). $\delta$
has been taken in the range of -0.02 to +0.02 in steps of 0.005.
 To start with, we calculate the equilibrium lattice 
parameter $($a$_{0}$$)$ as well as equilibrium volume 
$($V$_{0}$$)$, and the corresponding energy is considered 
as the equilibrium energy $($E$_{0}$$)$. Then strain 
 is applied to the system. Under this strained condition, 
the energy ($E$) is calculated and subsequent to that, the elastic 
constants are obtained from our calculations as discussed below.
The energies $\frac{E-E_0}{V_0}$ are plotted as a function of 
applied strain and fitted with a fourth order polynomial. 
The second order coefficient of the fit gives the elastic 
constants. The mechanical stability criteria for the cubic crystal are as follows: 
C$_{11}$ $>$ $0$, C$_{44}$ $>$ $0$, C$_{11}$$-$C$_{12}$ 
$>$ $0$ and C$_{11}$$+2$C$_{12}$ $>$ $0$. 
From the tables containing elastic constants we can see that first, 
second and fourth 
conditions are satisfied for all the materials listed here, 
but the third condition is not satisfied by some of the materials. 

{\bf Tetragonal shear constant (C$^\prime$)} - This is defined as 
0.5$\times$(C$_{11}$$-$C$_{12}$). A value of C$^\prime$ which is
close to zero or negative indicates that the material is mechanically 
unstable and prone to tetragonal distortion. 
It is clear from Table 3, that for Ni$_{2}$MnGa, as expected, 
the tetragonal shear constant is quite close to zero. For
 Ni$_{2}$MoGa the value of C$^\prime$ is found to be negative, which
indicates that it has a mechanically unstable cubic austenite
phase, which corroborates the result presented in Figure 5. It is interesting to note that 
Ni$_{2}$VGa has a value of C$^\prime$ close to that of 
Ni$_{2}$MnGa. From Figure 5, we observe that it has a flat region near
$c$/$a$ = 1, in the energy versus $c$/$a$ curve. Hence, in this case 
the ground state symmetry can not be properly ascertained 
  as is evident from our results. It is important to mention here
that the stoichiometric 
Ni$_{2}$MnSn is a material which is not known to undergo 
martensite transition and we find that the tetragonal shear 
constant has a slightly larger positive value of about 12 
compared to Ni$_{2}$MnGa. 

From Table 4, 
it is observed that two Co-based materials show a negative
or close to zero value for C$^\prime$. Out of these, Co$_{2}$NbSn
is known to exhibit non-cubic distortion.\cite{JPSJ58SF} Among the 
 materials, namely, Co$_{2}$MoGa, Co$_{2}$VGa, Co$_{2}$CrGa, 
Co$_{2}$CrGa, Co$_{2}$MnSn, Co$_{2}$TcGa, 
except the first alloy all others have a large positive value
for the $C^{\prime}$ constant.
It is also observed from Figure 5 that, for  
Co$_{2}$MoGa, there is a clear indication of the tetragonal phase
being the lowest energy state. This is not the case for the 
other materials. So the combined study of energetics and bulk 
mechanical properties of all the materials\cite{TRunpubl} 
indicates that the only two Co$_{2}BC$ materials which are likely 
to undergo tetragonal transition and to exhibit SMA property
are Co$_{2}$NbSn and Co$_{2}$MoGa. 

{\bf Inherent Crystalling Brittleness (ICB)} -  The calculated values of 
bulk modulus, $B$, have been listed in 
Tables 3 and 4 for Ni and Co-based materials, respectively. The 
isotropic shear modulus, $G$, is related to the resistance of 
the material to the plastic deformation. 
In literature, it has been shown\cite{hill} that the
value of $G$ lies in between the values of shear modulii given by 
formalisms of Voigt ($G_{V}$)\cite{voigt} and Reuss 
($G_{R}$)\cite{reuss}, which means $G$ = ($G_{V}$ + $G_{R}$)/2. 
As has been discused for austenite phase of Ni$_{2}$MnGa 
in our previous work (Ref.\onlinecite{JALCOM632TR}) 
 the experimental $G$ value is close 
to the calculated $G_{V}$ value while $G_{R}$ value remains largely 
underestimated. This occurs due to the small positive value of 
$C^{\prime}$. This particular aspect of similar FHAs, showing 
martensite transition, has already been discussed in detail in
 the literature.\cite{JALCOM632TR}. Following this observation,
we consider $G_{V}$ value as the value of shear modulus ($G$)
though it is generally considered to be the higher limit of the same.
Further, a simple and empirical relationship, given by 
Pugh\cite{PM45SFP}, proposes that the plastic property of a material 
is related to the ratio of the shear and bulk modulus of that 
particular material. 
A high value (greater than $\sim$ 0.57) of ratio of shear and bulk 
modulus, namely, $G$/$B$, is connected with the inherent crystalline 
brittleness of a bulk material. A value below this critical number 
phenomenologically signifies that the material's ICB is low. 
From Table 3, we find that 
the values for Ni-based materials are below this critical value and 
hence the ICB of these materials is low though the well-known 
FHA, Ni$_{2}$MnGa, has a somewhat higher value compared to the
 materials containing platinum in place of Ni.\cite{JALCOM632TR} 
We note that Ni$_{2}$MoGa show a negative value of C$^\prime$, as 
well as the lowest value of $G$/$B$ among all 
and hence it is expected to have a low ICB. 
So from energetics (Figure 5) and bulk mechanical points of view, 
it is a promising material, though not from magnetic point of view.
Table 4 lists the bulk mechanical properties for the Co-based FHAs studied 
here. The two materials which are likely to show a tetragonal ground state
(namely, Co$_{2}$NbSn and Co$_{2}$MoGa) are expected to exhibit 
ICB smaller or comparable to Ni$_{2}$MnGa ($G$/$B$ = 0.25 and 0.42, 
respectively). All the rest of the Co$_{2}BC$ alloys 
have $G$/$B$ values comparable to 
or larger than that of Co$_{2}$NbSn.

{\bf Cauchy Pressure} - We now focus on the value of  Cauchy Pressure,
 $C^{p}$, which is defined as $C^{p}$ =
$C_{12}$ - $C_{44}$. In Figure 9,
we plot the available data for $C^{p}$ versus $G_{V}$/$B$
calculated in case of some of the Ni and Co-based compounds.
We find that overall, there is a clear trend of
inverse (linear) relationship between $G_{V}$/$B$ and $C^{p}$. In the
literature also, it is observed that, the higher the $C^{p}$,
the lower the ratio $G$/$B$. Interestingly, this type of 
nearly-linear inverse relationship between the Cauchy pressure and 
the $G$/$B$ ratio seems to be a rather general observation as observed in 
the literature for various types of materials.\cite{JALCOM632TR,cpvsgbyb}

Finally, after analyzing the bulk mechanical as well as the magnetic 
properties and the energetics, out of all the materials studied, 
only two new materials, namely, Ni$_{2}$MoGa and Co$_{2}$MoGa,
emerge to be promising in 
terms of application as an SMA material. However, 
 due to the absence of any magnetic moment in Ni$_{2}$MoGa, 
this material is not expected to be suitable as an MSMA material. A low 
 $T_{C}$ indicates an absence of ferromagnetism in Co$_{2}$MoGa 
at room temperature as is observed in case of Co$_{2}$NbSn.

\subsection{Electronic Properties: Density of States}

\subsubsection{Analysis of Total and Atom-Projected Partial DOS}  

After discussing the energetics, magnetic and bulk mechanical 
property of the cubic austenite phase, we now
analyse the electronic property in terms of the total and partial
density of states of different atoms of various materials. 
We have carried out calculations on all the eletronically stable
materials\cite{TRunpubl} but here we concentrate on and present the 
results of the austenite phases of eight typical FM materials 
as discussed above. These materials are
Ni$_{2}$MnGa, Co$_{2}$MoGa, Co$_{2}$NbSn, which are to exhibit 
a tetragonal symmetry as well as Co$_{2}$VGa, Co$_{2}$CrGa,
 Co$_{2}$MnGa, Co$_{2}$MnSn and Ni$_{2}$MnSn, which are to possess a 
cubic symmetry, at the lowest possible temperature. 
We will discuss the total and atom-projected DOS of these systems 
in this section.

{\bf Total DOS}:

It is seen that the valence band width for all the materials is 
about the same, which is roughly about 6 eV (Figures 10 to 13). 
The two-peak structure in the DOS for both Ni and $C$ atoms 
indicating about
 substantial hybridization among these atoms is evident from 
Figure 10. A two-peak structure is observed at 
the Fermi level, this indicates a strong hybridization between the Ga 
 and Ni atoms. This covalent interaction between the Ga $4p$ and 
Ni $3d$ minority electrons plays a crucial role in the stability.
We note here that the overlapping and the two-peak structure of the DOS 
is prominent in case of $A$ and $C$ atoms of Co$_{2}$ZrGa, which has 
a highly negative formation energy (Figure 4).
On the other hand, in Co$_{2}$CrGa, the two-peak structure 
 and the overlapping of DOS for both the $A$ and $C$ atoms 
are not quite substantial and the formation 
energy is low as well (Figure 4).\cite{TRunpubl} Further, 
Figure 13 depicts the DOS for $B$ = Y and Zr. These plots indicate 
a lowering of binding energy in the $B$ = Zr material compared to 
the $B$ = Y case. This corroborates the trend of the formation energy 
values of these materials. It is to be noted that the 
contribution from the $A$ atom plays a crucial role in this. 
Zayak et al\cite{PRB72ATZ} have earlier shown that the
stability of the Ni$_{2}$MnGa type Heusler alloys is 
closely related to the minority DOS at the Fermi level as has been argued 
in other cases as well.\cite{PRB84CL} 

There is a large exchange splitting observed for 
systems which have Mn as the $B$ atom. From Figures 10 to 12,
we observe that for $A_{2}BC$ systems ($A$ = Ni, Co; $B$ = Mn; $C$ = 
Ga; Sn), the occupied DOS of the $B$ atom is dominated by the majority 
spin whereas the unoccupied DOS is dominated by the minority spin. 
For Ni$_{2}$MnGa and Ni$_{2}$MnSn, the majority DOS of the Mn atom 
is centred around -1.2 eV and -1.5 eV, whereas the minority DOS of 
the same atom is centred around 1.5 eV and 1.3 eV for the 
respective systems. For Co$_{2}$MnGa and Co$_{2}$MnSn, the position 
of the occupied majority spin DOS for Mn atom is at about -0.7 eV and -1.1 
eV, respectively, while the position of the unoccupied minority 
DOS peak is at about 1.8 eV and 1.6 eV. For Co$_{2}$CrGa also, we observe 
a large separation between the occupied majority DOS peak (at about -0.1 eV) 
of Cr atom and unoccupied minority DOS peak (at about 1.7 eV).  
Next, we analyze the partial DOS of few of the important 
$A$, $B$ atoms, to understand the nature of DOS close to the Fermi level.

{\bf Partial DOS} : 

{\it Ni Atom} - The DOS in case of the two Ni-based 
alloys are similar (Figure 10). However, since Sn atom contains
 one extra velence electron 
compared to the Ga atom in the $C$ position, the peak positions of 
the total DOS of the Mn atom are shifted towards lower energy in case 
of the materials with $C$ = Sn. 
At this point, it is worth-mentioning that it has 
already been discussed in the literature that a rigid band model is 
a suitable model to understand the trends when the $C$ atom is 
changed.\cite{JPCM11AA} 
We further observe that a similar situation is seen to arise when 
 the $A$ atom is changed from Ni to Co, which is discussed below.

{\it Co Atom} - 
Co has one velence electron less than Ni. Hence, a larger contribution
of Co-derived levels compared to Ni-derived levels in the unoccupied 
part of the respective DOS is expected. Figures 11 to 13 depict this.
When DOS of Ni$_{2}$Mn$C$a is compared with Co$_{2}$Mn$C$, it is clearly 
evident (Figures 10, 11 and 12).
Among the materials with $C$ = Ga and $A$ = Co, only Co$_{2}$MoGa
 is a material which is likely
to show a martensite transition (Figure 5). It is seen that it
has the first unoccupied DOS peak very close to the $E_{F}$ (Figure 11). 
Among the materials with $C$ = Sn and $A$ = Co, only Co$_{2}$NbSn
is known to be prone to distortion\cite{JPSJ58SF} and it has an unoccupied
DOS peak close to $E_{F}$ as well.
 When we analyze the DOS of Co$_{2}$MnGa (Figure 11) and 
Co$_{2}$MnSn (Figure 12), which do not show the tendency of a 
tetragonal distortion as well as are known to possess high spin
polarization at the Fermi level, we observe that the first 
unoccupied DOS is further away from $E_{F}$ compared to the
materials which are prone to tetragonal distortion, namely Co$_{2}$MoGa
(Figure 11) and Co$_{2}$NbSn (Figure 12). 

{\it Mn Atom} - There are four out of eight
materials which contain Mn atom in the $B$ position. 
When we compare the total DOS of the Mn atom at the $B$ position, 
in all the four materials considered here, it is clearly seen that 
majority of the 
DOS of the down spin occupies the unoccupied region above the Fermi 
level, while the up spin electrons primarily 
have negative binding energies. 
As opposed to the down spin DOS, which has one major peak in all the
four cases, the up spin electrons typically occupy 
two energy ranges, one around 1 and one around 3 eV below $E_{F}$. 
Due to one extra electron in Sn atom
compared to the Ga atom in the $C$ position, the peak positions of 
the total DOS of the Mn atom are shifted towards lower energy in case 
of the materials with $C$ = Sn. To elaborate, first we compare the 
DOS of the up spin of Mn in the four alloys Ni$_{2}$MnGa, 
Ni$_{2}$MnSn, Co$_{2}$MnGa and Co$_{2}$MnSn in the cubic phase.
While DOS of Mn atom in the unoccupied part peaks at about 1.5 eV
in case of $C$ = Ga, it peaks around 1.2 eV when $C$ = Sn. 
The corresponding peak positions for Co$_{2}$MnGa and Co$_{2}$MnSn 
are at about 1.8 and 1.6 eV, respectively.
In case of the up spin DOS, there are two ranges of predominant
DOS in all the four materials. For the first such range, which is
closer to the Fermi energy, the peak positions are at about
-1.3, -1.5, -0.7 and -1.1 for Ni$_{2}$MnGa, Ni$_{2}$MnSn, Co$_{2}$MnGa
and Co$_{2}$MnSn, respectively. For the range which is at a much 
higher binding energy, the peak positions are at about
-3.2, -3.2, -2.5 (also one slightly weaker one at -2.8)
 and -2.5 (also one slightly weaker one at -2.8) eV
 for Ni$_{2}$MnGa, Ni$_{2}$MnSn, Co$_{2}$MnGa
and Co$_{2}$MnSn, respectively. It is to be noted that the peaks
of the DOS are not sharp but broad ones, with shoulders on either
or one of the sides.

\subsubsection{Electronic Stability of the Tetragonal phase from DOS}  

After discussing the electronic                      
property of the cubic austenite phase, we now
analyse the electronic property in terms of the
density of states of different materials as a function of $c$/$a$. 
We concentrate on the eight typical FM materials as discussed above. 
A tetragonal distortion has been imposed on all these eight materials. 
To highlight the difference between the two symmetries, we will 
concentrate on the detailed results of cases with 
$c$/$a$ = 1, 1.05 and 1.10.
The aim is to understand the electronic stability or instability of 
the tetragonal phase of these compounds from the DOS results.   

{\bf Ni$_{2}$MnGa versus Ni$_{2}$MnSn} :

 Figure 14 contains 
the density of states of     
the cubic and tetragonal phases, with $c$/$a$ varying from 1 to 1.10
in steps of 0.05 for materials Ni$_{2}$MnGa and Ni$_{2}$MnSn. 
First we will analyse Figure 14 for the cubic phase. We observe that
there is a peak at around -0.2 eV for Ni$_{2}$MnGa and at around -0.5
eV for Ni$_{2}$MnSn, respectively. As has been established in the
literature, this peak, which is close to the $E_{F}$, has negative
binding energy. This peak is derived from the electrons of the Ni atoms 
with down spin having $e_{g}$ symmetry and is known to 
 play a crucial role in the stabilization of the tetragonal 
phase in case on Ni$_{2}$MnGa.\cite{JPCM11AA} The density of states 
of the down spin electrons with $t_{2g}$ symmetry of these $A$ atoms 
corresponds to the peaks with reasonably higher binding energy. 
This is the case for both the materials. On the other hand, detailed
investigation suggests that the $B$ atom = Mn has negligible
 contribution 
near the Fermi level; both for the up and for the down spin. The 
up spin electrons of $A$ atom 
also do not significantly contribute to the DOS at around -0.2 
and -0.5 eV for Ni$_{2}$MnGa and Ni$_{2}$MnSn, respectively.

 As $c$/$a$ 
increases in case of these materials, there are some
systematic changes in the density of states, clearly visible from the 
lower panels of the Figures 14 to 17.
For Ni$_{2}$MnGa, it is seen that the peak near the Fermi level, at
about -0.2 eV, 
derived from the down spin DOS, has been split into two peaks.
This has been observed and argued about in detail 
in the literature.\cite{JPCM11AA,JPCM11PJB,SSC18JCS} 
As a result of tetragonal
distortion, the degeneracy of the sub-bands near
the Fermi level is lifted. As a consequence, a redistribution of
the density of states of the $3d$ electrons and in turn a reduction
of free energy occurs. This is the so-called band Jahn-Teller effect 
which is known to result in the lowering of energy under tetragonal
distortion in many FHAs including 
Ni$_{2}$MnGa.\cite{JPCM11AA,JPCM11PJB} 
For Ni$_{2}$MnSn as well, it is seen that the most prominent change
being the spilitting of the peak at about -0.5 eV\cite{JPCM11AA} 
upon the tetragonal distortion as seen from Figure 14. However, it is 
well-known that stoichiometric Ni$_{2}$MnSn is not expected to
have tetragonal ground state.
It has been argued in the literature that 
the band Jahn-Teller effect is sensitive to the DOS at the $E_{F}$
in the cubic phase. The closeness of the degenarate peak for 
Ni$_{2}$MnGa (at about -0.2 eV with respect to $E_{F}$) over 
Ni$_{2}$MnSn (at about -0.5 eV below $E_{F}$) is an indication 
of the possibility of tetragonal distortion in the
former.\cite{JPCM11AA,JPCM11PJB}
We find that the density of states at $E_{F}$ in case of cubic phase
of Ni$_{2}$MnGa is relatively more in comparison to Ni$_{2}$MnSn, 
which is evident from the relative position of the $e_{g}$ peak 
near the Fermi level in the two materials (Figure 14).

{\bf Co$_{2}$MoGa versus Co$_{2}$MnGa} :

Figure 15 gives the plot of DOS with different $c$/$a$ values. 
We note that in the literature the stability of the martensite 
phase for a Co-based system (Co$_{2}$NiGa) has been explained to 
be due to the lowering of energy of the system under the tetragonal 
deformation compared to the cubic structure.\cite{ActaMat58RA} 
In case of Co$_{2}$MoGa there is a large peak in the minority DOS 
just above the Fermi level (at about +0.3 eV with respect to $E_{F}$).
 Detailed analysis
 shows that this peak at the minority DOS has contributions from 
all three atoms i.e. (Co, Mo, Ga), but the major contribution comes from 
the $e_{g}$ levels of $3d$ electrons of Co. We find that these Co $e_{g}$ 
levels have a 
 major role in the stabilization of the tetragonal phase, similar to 
the Ni $e_{g}$ levels in case of Ni$_{2}$MnGa. 
Hence, in case of Co$_{2}$MoGa also, band Jahn-Teller distortion 
plays a significant role. The down spin DOS close to the Fermi level is 
high, 
which leads to the instability of the cubic phase of this material unlike
Co$_{2}$MnGa. In case of the latter material, 
 the minority DOS almost vanishes at $E_{F}$.
Further, here, the $e_{g}$ levels of 
Co atom are located farther away from the Fermi level (at about +1.0 eV 
with respect to $E_{F}$). 
For both the materials $B$ atom does not contribute to 
the minority DOS at the Fermi level. However, primarily the 
$B$ atom only contributes significantly to the large DOS of the up spin
electrons at the Fermi level for Co$_{2}$MoGa. The 
 peak positions of the DOS in the unoccupied part of the energy 
for Co$_{2}$MoGa and Co$_{2}$MnGa are different due to the hybridization
of Co atom with the Mo and Mn atoms, respectively. This led to the
difference in the electronic characters of these two materials.

{\bf Co$_{2}$NbSn versus Co$_{2}$MnSn} : 

Figure 16 gives the plot of DOS with different $c$/$a$ values. 
As in case of Ni$_{2}$MnGa, for Co$_{2}$NbSn also we can find 
same type of evolution of density of states as a function of 
$c$/$a$ and Co $3d$ $e_g$ states play the key role in the tetragonal 
transition. In austenite phase of Co$_{2}$NbSn the Co $3d$ 
$e_g$ peak is at just above Fermi level (at about +0.05 eV). Under 
tetragonal distortion this peak is split into two: one part 
being above the Fermi level and another one being below the 
Fermi level. This splitting lowers the energy of the system
and the tetragonal phase tends to be the ground state structure 
compared to the cubic structure. 
In the literature\cite{JPSJ58SF}, it has been observed that at the 
Fermi energy, the contribution to DOS mainly comes from the $3d$ bands 
of Co and Nb. We observe from Figure 12 that Sn atoms also contribute. 
In the cubic phase Co atom has a single large peak just 
above the $E_{F}$. But under tetragonal distortion, this 
single peak is split into two and the energy of the tetragonal phase 
is lower compared to the cubic phase. 
It has been observed\cite{JPSJ58SF} that 
the band Jahn-Teller distortion is the cause of the structural transition.
As seen from Figures 12 and 16, for Co$_{2}$MnSn, the peak due to Co 
$3d$ $e_g$ levels is located at 
a higher energy (at about +0.8 eV) compared to Co$_{2}$NbSn. 
After the application of the tetragonal distortion this single peak 
is split into two (Figure 16). But in this case after splitting both the 
peaks lie above $E_{F}$ which does not yield to lowering of energy.

{\bf Co$_{2}$CrGa versus Co$_{2}$VGa} : 

 From Figure 17,
 it is observed that the single peak of the Co $3d$ $e_g$ above $E_{F}$ 
is split for Co$_{2}$VGa. However, after splitting both the peaks lie 
above $E_{F}$. So there is no lowering of energy of the system possible 
under tetragonal deformation. For Co$_{2}$CrGa, the Co $3d$ $e_g$ 
single peak of cubic phase is located at an even higher (positive) energy
 with respect to the Fermi level. 
In this case also lowering of energy under tetragonal deformation is not 
possible, which is consistent with the results presented in Figure 5.

Finally, we find that in all the materials,
 as a result of tetragonal distortion,
the degeneracy is lifted for the $3d$ sub-bands in the minority
spin channel of $A$ atoms, present closest to the Fermi level.
Subsequently, as a result of this band Jahn-Teller effect, 
a redistribution 
of the density of states of these $3d$ electrons occurs.
In all the materials, which favor a tetragonal deformation, 
a substantial density of states very close to the Fermi level has
been observed. In other words, the band Jahn-Teller effect is 
found to be, as expected, quite 
sensitive to the DOS at or close to the $E_{F}$ in the cubic phase. 
As a result of the redistribution of the DOS, 
under tetragonal distortion, due to closeness of peak in DOS 
to the $E_{F}$, the 
  energy gets lowered in these materials. 
Consequently, the possibility
of martensite transition is found to be high. Further, 
for these materials, a negative or very close to zero value of 
tetragonal shear constant, C$^\prime$, has also been observed, 
as is expected from the literature. For all the 
other materials, under tetragonal distortion, the splitting of the 
$3d$ minority spin levels is observed as well but  
the peak is away from the $E_{F}$ resulting in 
 a reduced density of states at $E_{F}$. 
Therefore, the lowering of free energy is not possible, which renders the
tetragonal transition unlikely.
The observation regarding the cubic ground state for these materials 
is further corroborated by the relatively 
large and positive values of $C\prime$ for all these materials.

\section{Summary and Conclusion}

It has been of particular interest that out of all the Ni and 
Co-based full-Heusler
alloys studied so far in the literature, only some materials undergo 
the martensite transition and these generally show the 
technologically important magnetic shape memory alloy property. 
These full Heusler alloys in general are found to be metallic in
nature. On the other hand, it has been observed that there is another
group of FHAs which are half-metallic-like in nature,
with a much reduced density of states at the Fermi level
in case of one of the spin channels and these materials generally
do not exhibit MSMA property. It has been observed earlier  
that while most of the Ni-based FHAs show MSMA property, many of the
Co-based FHAs exhibit large spin-polarization at the Fermi level.

Therefore, in this paper, using first-principles calculations 
based on density functional theory, we study in detail the bulk mechanical, magnetic and electronic properties of a series of Ni and Co-based full Heusler alloys, namely, Ni$_{2}BC$ and Co$_{2}BC$
($B$ = Sc, Ti, V, Cr and Mn as well as Y, Zr, Nb, Mo and Tc;
$C$ = Ga and Sn). After establishing the electronic stability from the 
formation energy and subsequent full geometry optimization, we carry 
out the calculation of different properties to probe and understand, 
the similarities and differences in the properties 
of these materials. We analyze the data in detail to 
see if among these materials there is any FHA which has a tendency to
undergo a tetragonal transition and at the same time possesses
a high spin polarization at the Fermi level.

Out of all the electronically stable compounds of the total forty Ni and 
Co-based materials, most of the Ni-based materials are expected
to show a non-magnetic ground state. On the other hand, 
Ni$_{2}$MnGa and Ni$_{2}$MnSn
as well as all the Co-based materials are ferromagnetic in nature.
Further, from the Heisenberg exchange interaction parameters, it
is seen that the materials exhibit similar nature in terms
of the relative contributions of the direct and RKKY-type nature
of the magnetic interactions. The trend of the calculated values of 
Curie temperature for various materials, obtained from the 
$J_{ij}$ parameters, matches reasonably well with the literature
wherever data are available.
From the point of view of bulk mechanical properties, the values 
of tetragonal shear constant show consistent trend: high positive for 
materials not prone to
tetragonal transition and low or negative for others. A general trend
of nearly-linear inverse relationship between the Cauchy pressure and 
the $G$/$B$ ratio is predicted for both the Ni and Co-based materials.

It is observed that the Ni-based materials
are typically metallic in nature. However, 
all the Co-based alloys exhibit a significant spin polarization
at the Fermi level. Most 
of the Ni-based materials have a $3d$ band of the minority spin of the $A$ 
atoms close to and below the $E_{F}$. On the other hand, the peak position
of the same band is above the $E_{F}$ for the 
 Co-based materials. We observe 
that, in both the cases of Ni and Co-based materials, these $3d$ 
levels play an important role in deciding the ground state. 
Further, the replacements of the $A$, $B$, $C$ 
sites of the $A_{2}BC$ materials by different atoms, indicate that 
in general a rigid band model explains the differences in the electronic 
structure of both the Ni and Co-based materials to a large extent. 
This model along with the hybridization between atoms, 
further supports the results of partial and total moments of 
these systems.
The relation between the closeness of the peak corresponding to the 
$e_{g}$ levels of the $3d$ down spin electrons of the $A$ atom to the 
$E_{F}$ and the tendency of lowering of energy upon tetragonal 
distortion is consistent across all the Ni and Co-based materials.

Finally, from our study on the two categories of materials, it is 
clear that out of all the materials which we study here, only 
four FHAs show a tendency of undergoing martensite transition. Out
of these four materials, which have a conventional Heusler alloy 
structure and exhibit a clear possibility of finding
a tetragonal phase as their ground state, three of them, namely,
Ni$_{2}$MnGa, Ni$_{2}$MoGa and Co$_{2}$NbSn have a metallic
 nature as is observed in case of majority of the MSMA material; on 
the other hand, from our calculations, Co$_{2}$MoGa is expected to 
emerge as a {\it shape memory alloy with high spin polarization at
the Fermi level}. 
This interesting finding awaits a suitable experimental validation.

\section{Acknowledgments}  
Authors thank P. D. Gupta, P. A. Naik and G. S. Lodha for facilities
and encouragement throughout the work. The scientific computing group,
 computer centre of RRCAT, Indore and P. Thander are thanked for
help in installing and and support in running the codes. 
S. R. Barman, A. Arya, S. B. Roy and C. Kamal are thanked for
 discussion. TR thanks HBNI, RRCAT for financial support.

{}

\clearpage
\pagebreak

\begin{figure}
\centering
\includegraphics{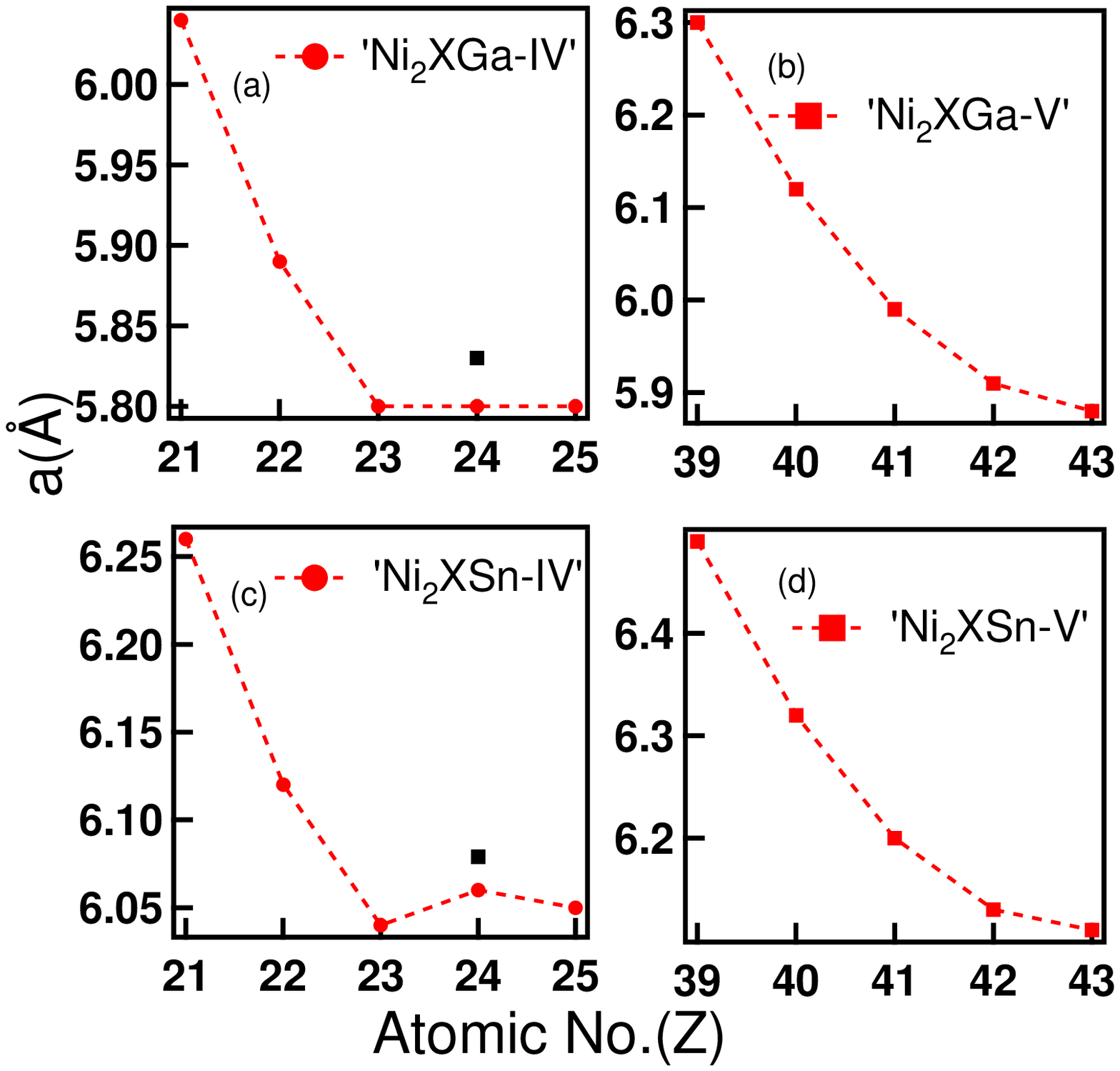}
\caption{
Variation of lattice parameter as a function of $Z$ of $B$ elements 
for Ni$_{2}BC$ alloy ($C$ = Ga, Sn); X=$B$ atoms being first five 
transition metal elements of period IV (left panel) and V (right panel). 
}
\label{fig:1}
\end{figure}

\clearpage
\pagebreak

\begin{figure}
\centering
\includegraphics{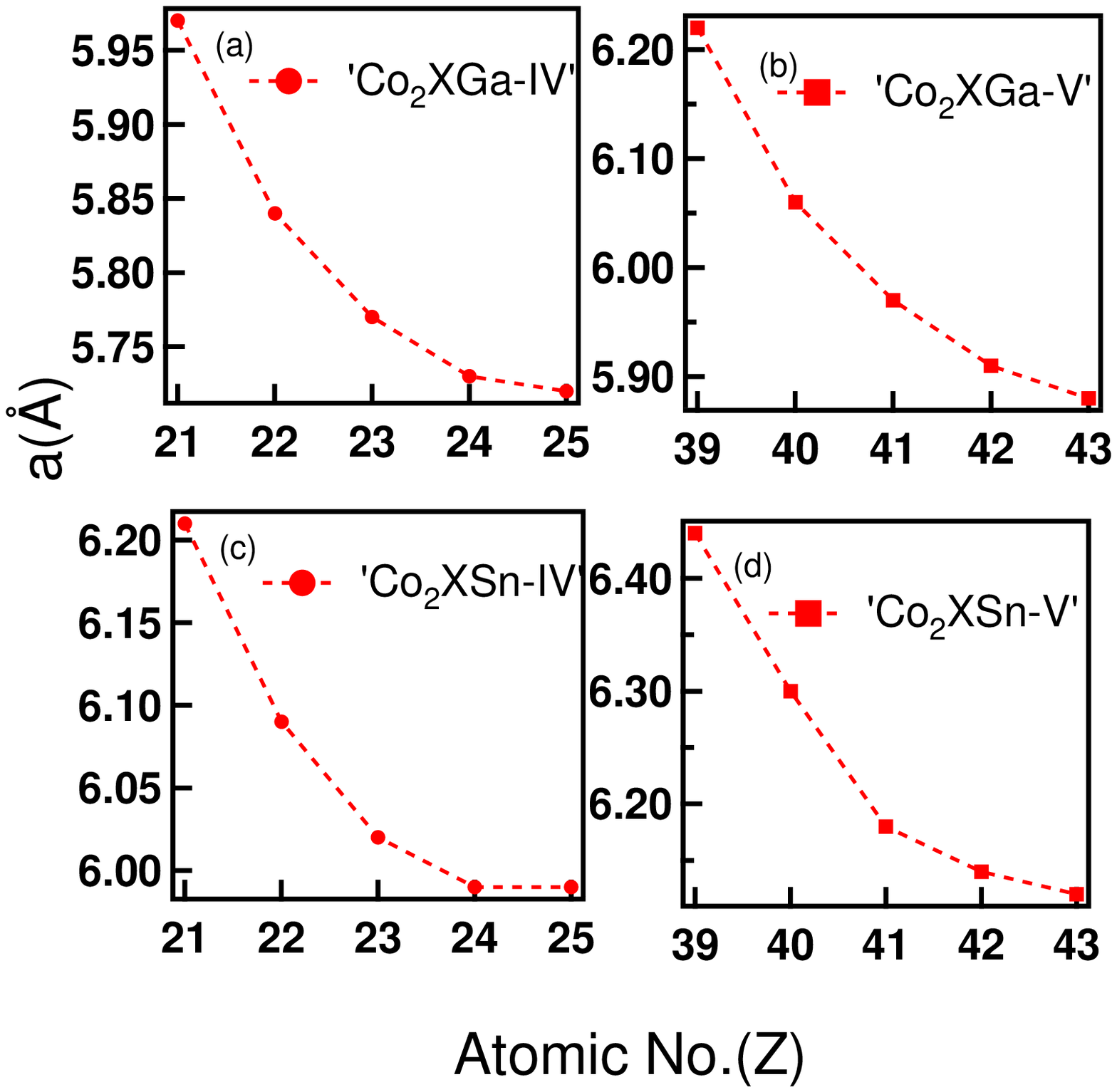}
\caption{
Variation of lattice parameter as a function of $Z$ of $B$ elements 
for Co$_{2}BC$ alloy ($C$ = Ga, Sn); X=$B$ atoms being first five 
transition metal elements of period IV (left panel) and V (right panel). 
}
\label{fig:2}
\end{figure}

\clearpage
\pagebreak

\begin{figure}
\centering
\includegraphics{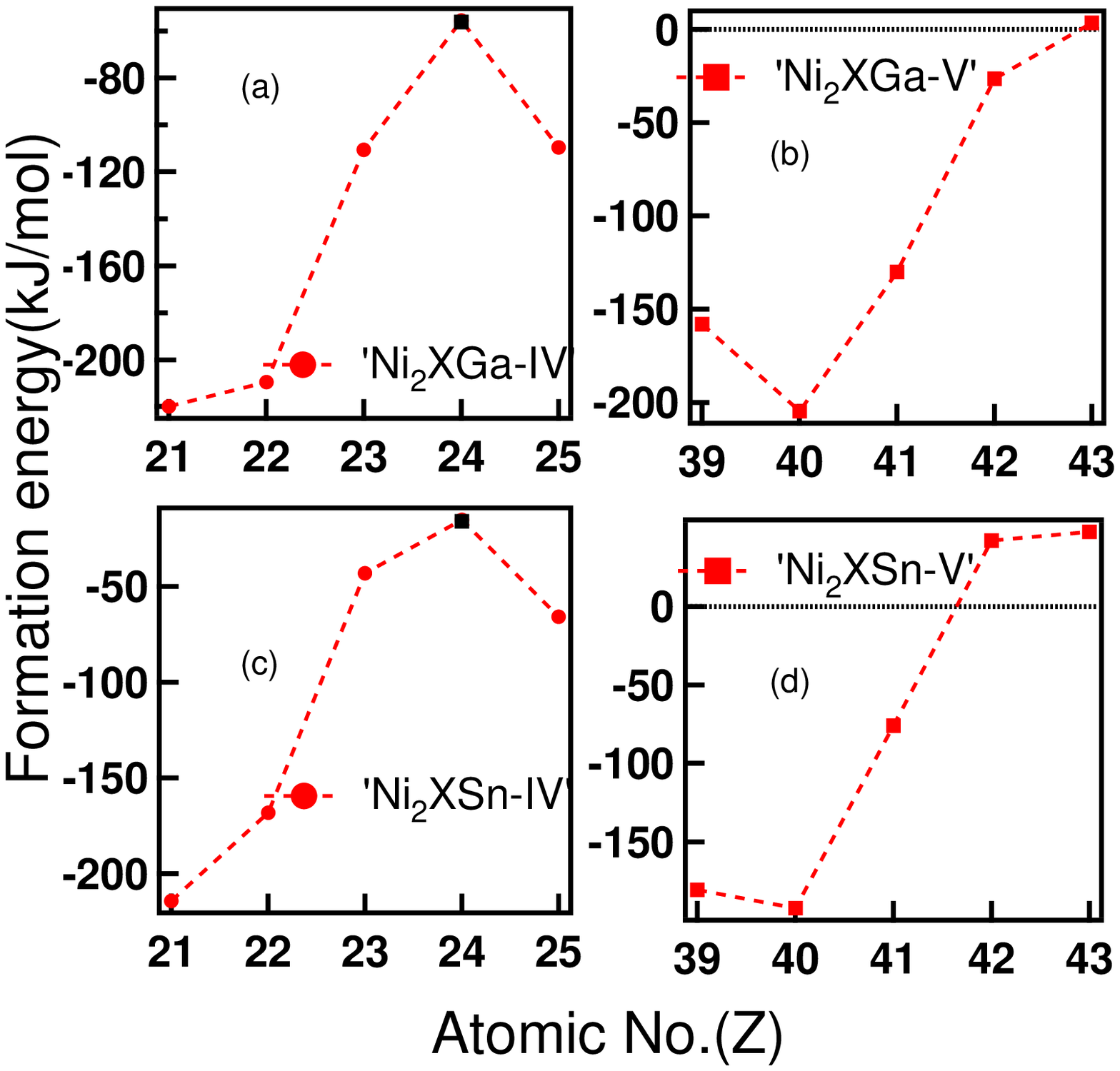}
\caption{
Variation of formation energy as a function of $Z$ of $B$ elements 
for Ni$_{2}BC$ alloy ($C$ = Ga, Sn); X=$B$ atoms being first five 
transition metal elements of period IV (left panel) and V (right panel). 
}
\label{fig:3}
\end{figure}

\clearpage
\pagebreak

\begin{figure}
\centering
\includegraphics{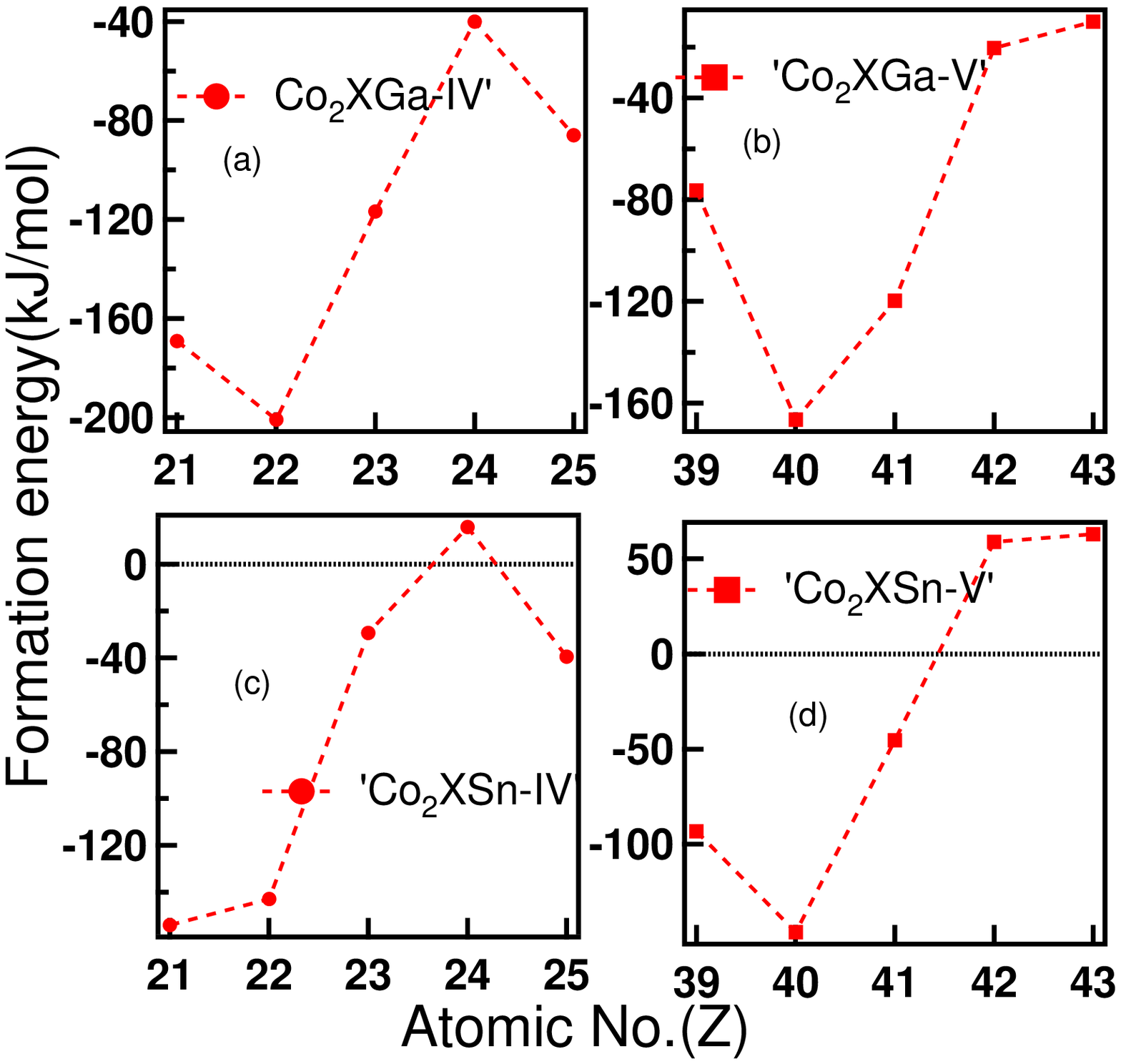}
\caption{
Variation of formation energy as a function of $Z$ of $B$ elements 
for Co$_{2}BC$ alloy ($C$ = Ga, Sn); X=$B$ atoms being first five 
transition metal elements of period IV (left panel) and V (right panel). 
}
\label{fig:4}
\end{figure}

\clearpage
\pagebreak

\begin{figure}
\centering
\includegraphics{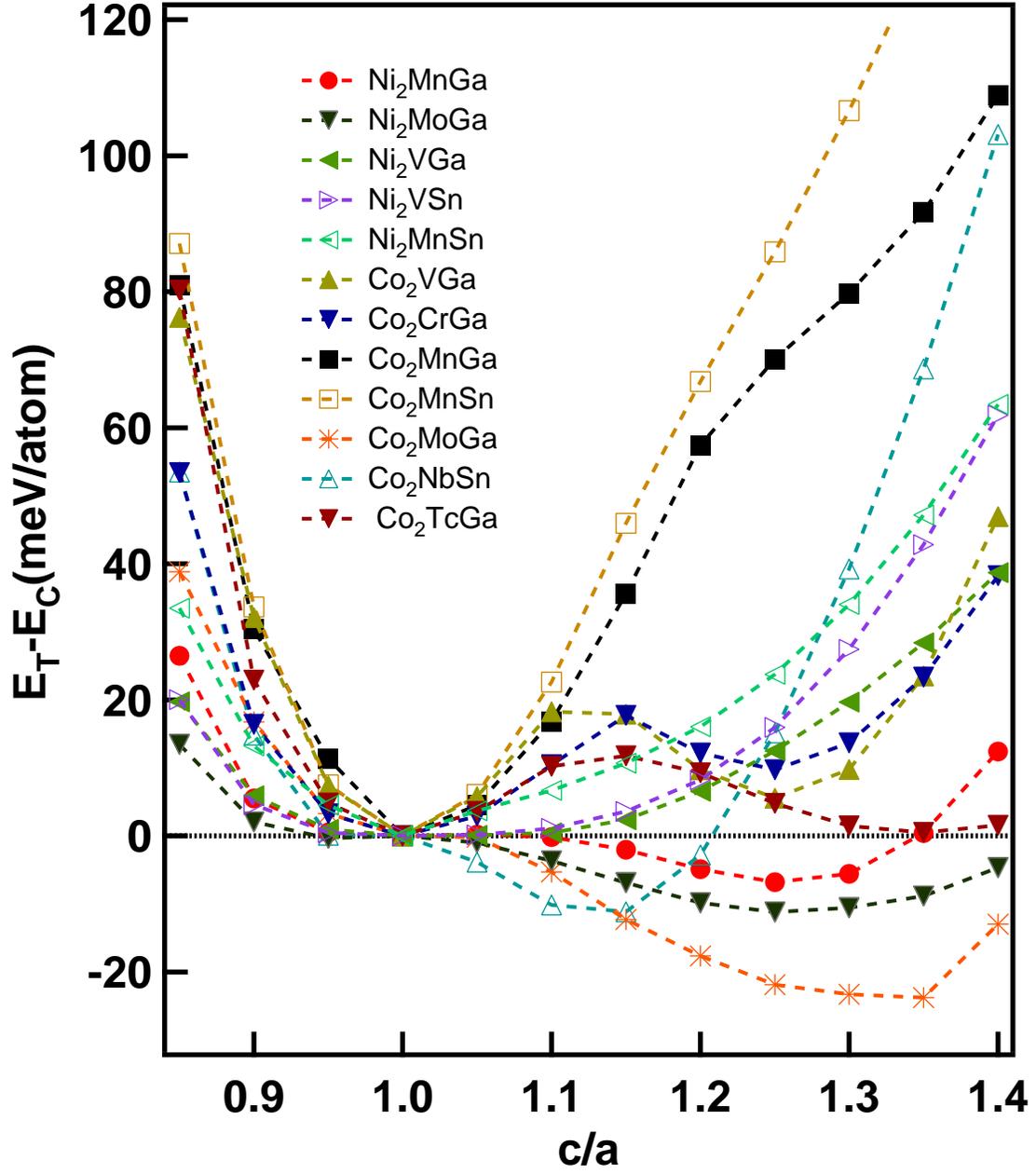}
\caption{
Energy difference between the crystal structures with 
tetragonal (T) and cubic (C) symmetries, of some typical materials
represented as $E_{T} - E_{C}$ (in units of meV per atom), 
as a function of the ratio of lattice constants $c$ and $a$. The
energies have been normalized with respect to the energy of the
respective cubic austenite phase of each material.
}
\label{fig:5}
\end{figure}

\clearpage
\pagebreak

\begin{figure}
\centering
\includegraphics{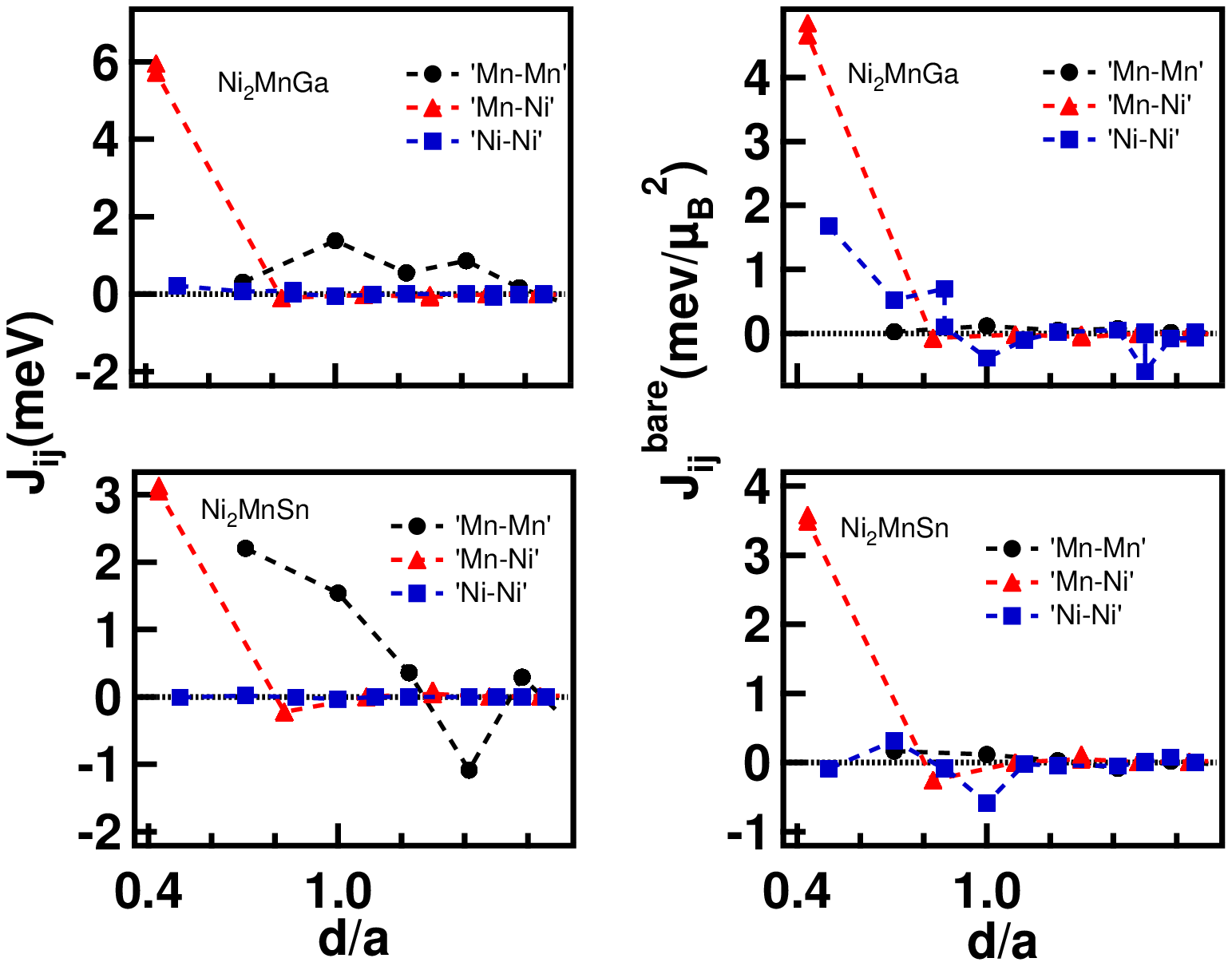}
\caption{
$J_{ij}$ parameters between different atoms of Ni$_{2}$MnGa and 
Ni$_{2}$MnSn as a function of distance between the atoms $i$ and
$j$ (normalized with respect to the respective lattice constant).
}
\label{fig:6}
\end{figure}

\clearpage
\pagebreak

\begin{figure}
\centering
\includegraphics{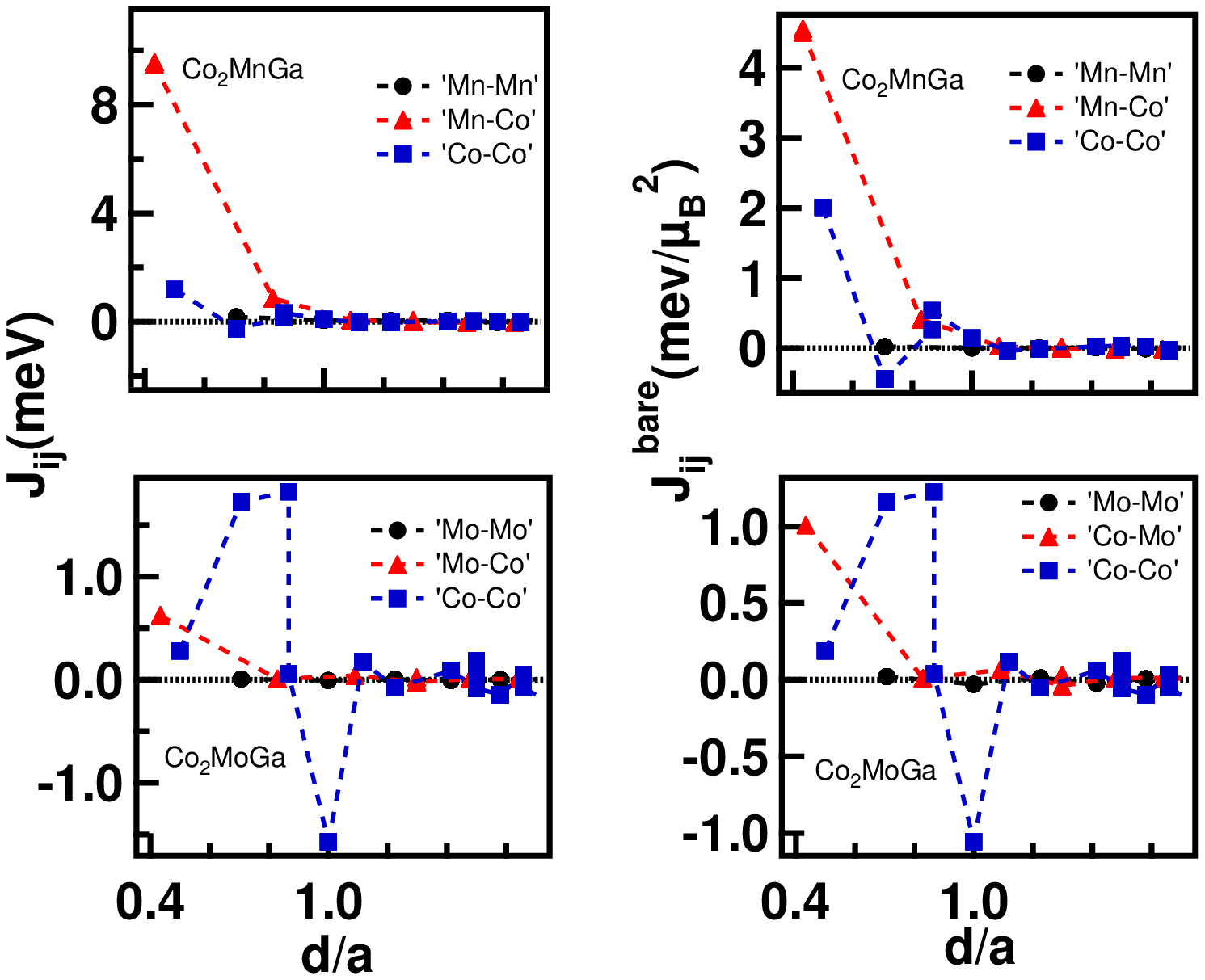}
\caption{
$J_{ij}$ parameters between different atoms of Co$_{2}$MnGa and 
Co$_{2}$MoGa as a function of distance between the atoms $i$ and
$j$ (normalized with respect to the respective lattice constant).
}
\label{fig:7}
\end{figure}

\clearpage
\pagebreak

\begin{figure}
\centering
\includegraphics{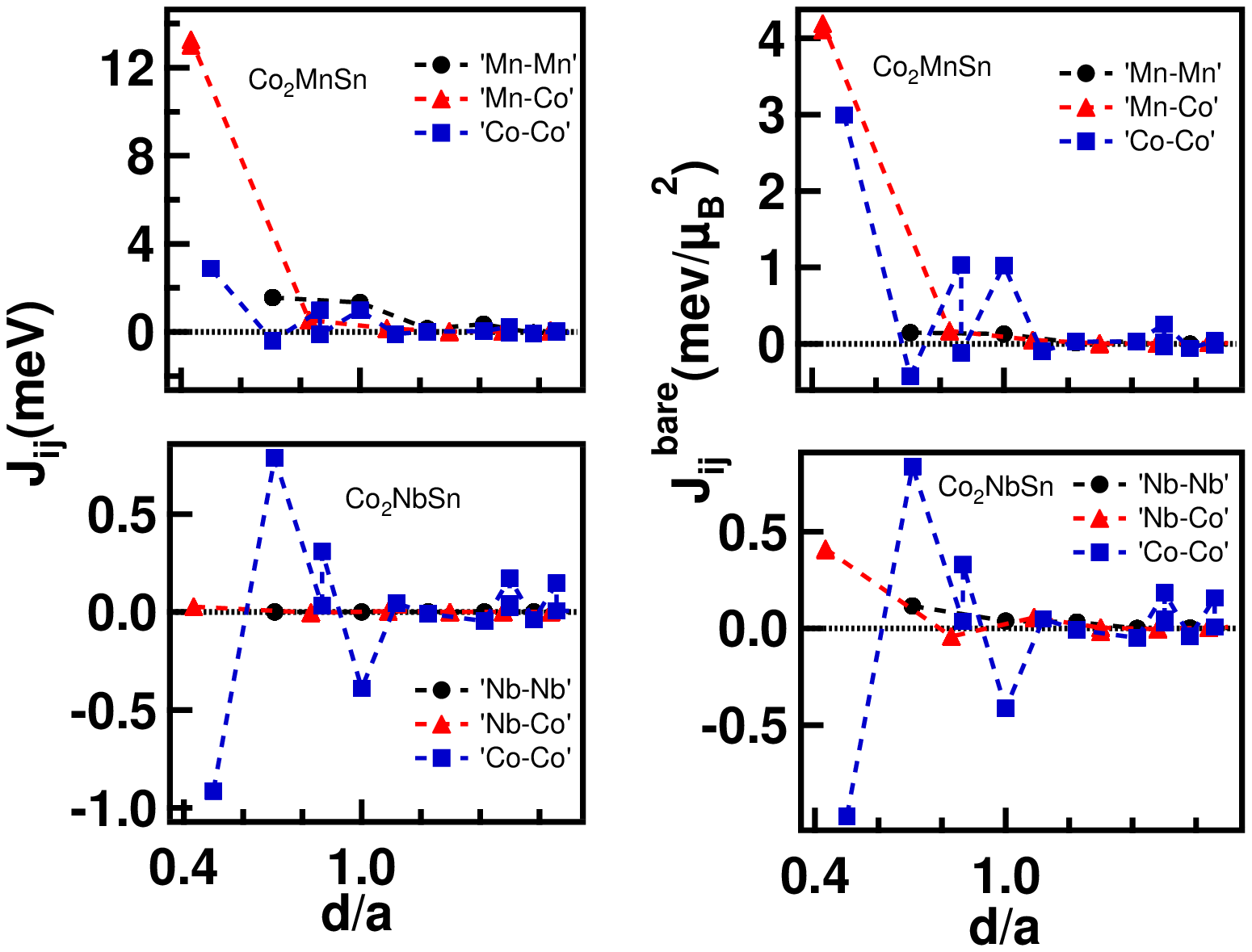}
\caption{
$J_{ij}$ parameters between different atoms of Co$_{2}$MnSn and 
Co$_{2}$NbSn as a function of distance between the atoms $i$ and
$j$ (normalized with respect to the respective lattice constant).
}
\label{fig:8}
\end{figure}

\clearpage
\pagebreak

\begin{figure}
\centering
\includegraphics{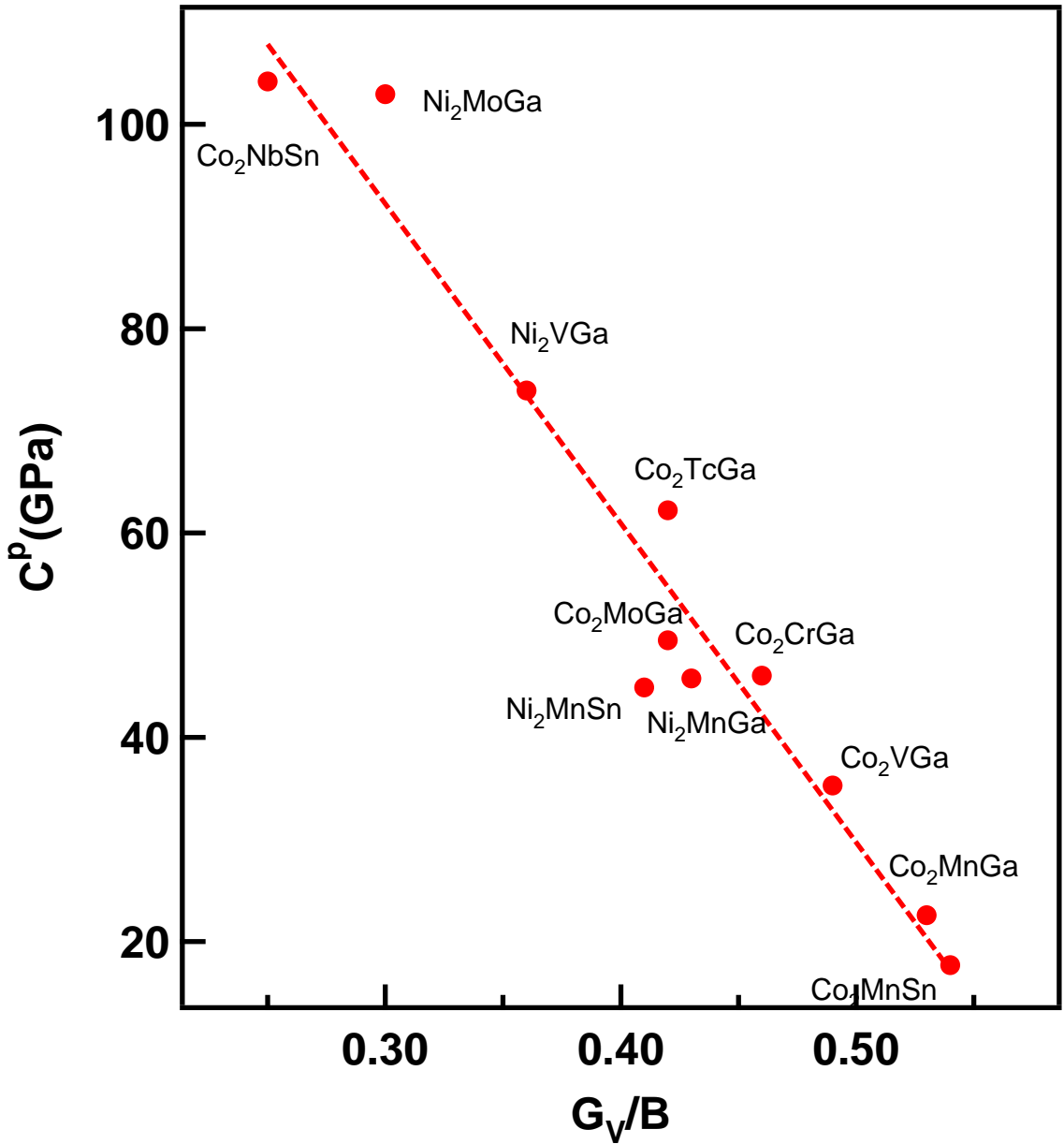}
\caption{Cauchy pressure, $C^{p}$, versus $G_{V}$/$B$; a linear 
fitting of all the data is carried out and shown here. An inverse 
linear-type relation is seen to exist between the two parameters 
(see text). 
}
\label{fig:9}
\end{figure}

\clearpage
\pagebreak

\begin{figure}
\centering
\includegraphics{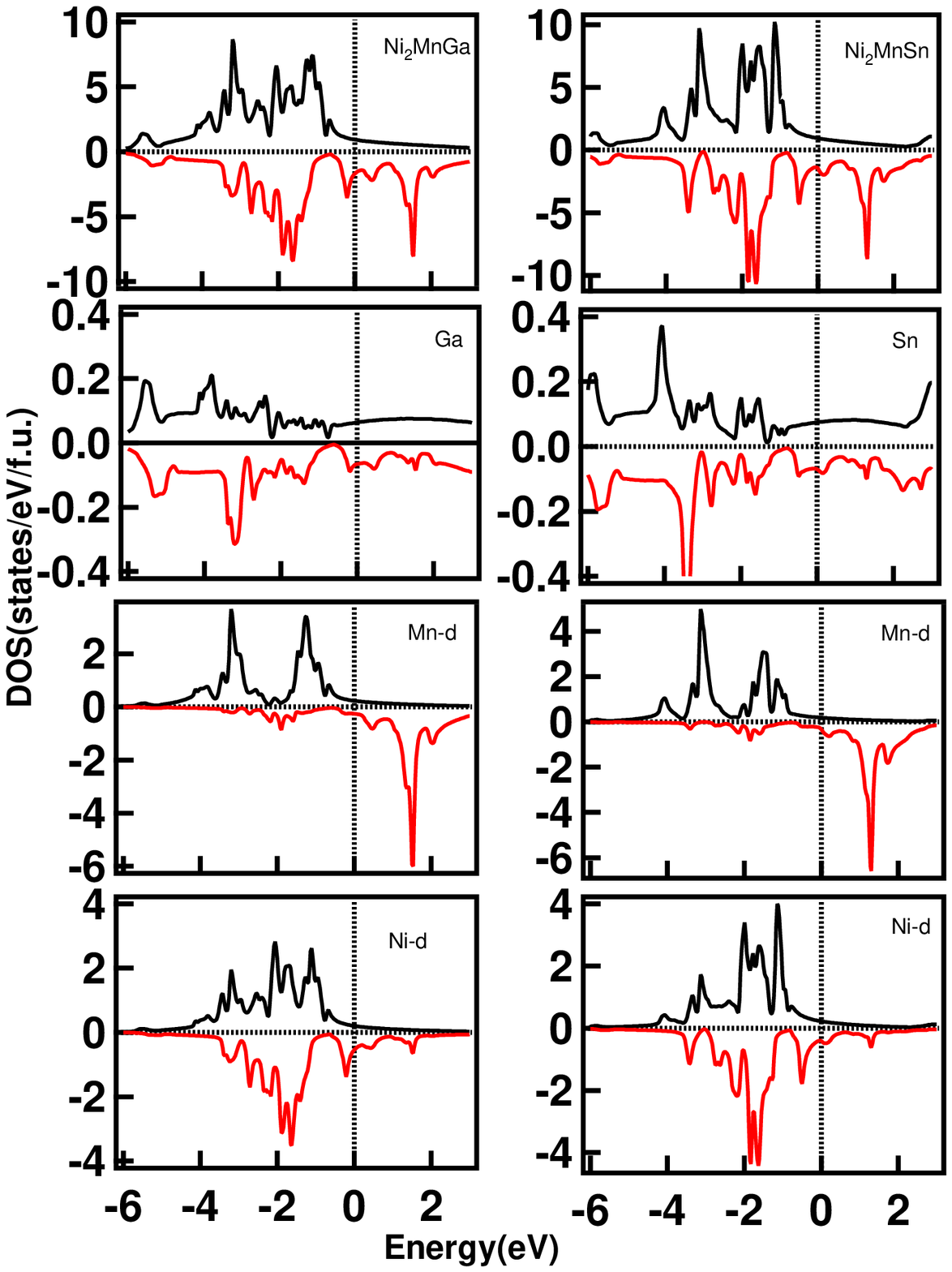}
\caption{
The left 
and right set of panels depict the density of states of Ni$_{2}$MnGa 
and Ni$_{2}$MnSn materials, respectively. 
From top to bottom panel, first the total density of states
as a function of energy has been plotted. 
Next panel shows the partial density of states of the Ni atom. 
Partial density of states of the Mn atom and the $C$ atom are 
shown in the third and the fourth panels. 
The DOS of the $C$ atom is multiplied by a factor of 10.
The Fermi level is at 0 eV.
}
\label{fig:10}
\end{figure}

\clearpage
\pagebreak

\begin{figure}
\centering
\includegraphics{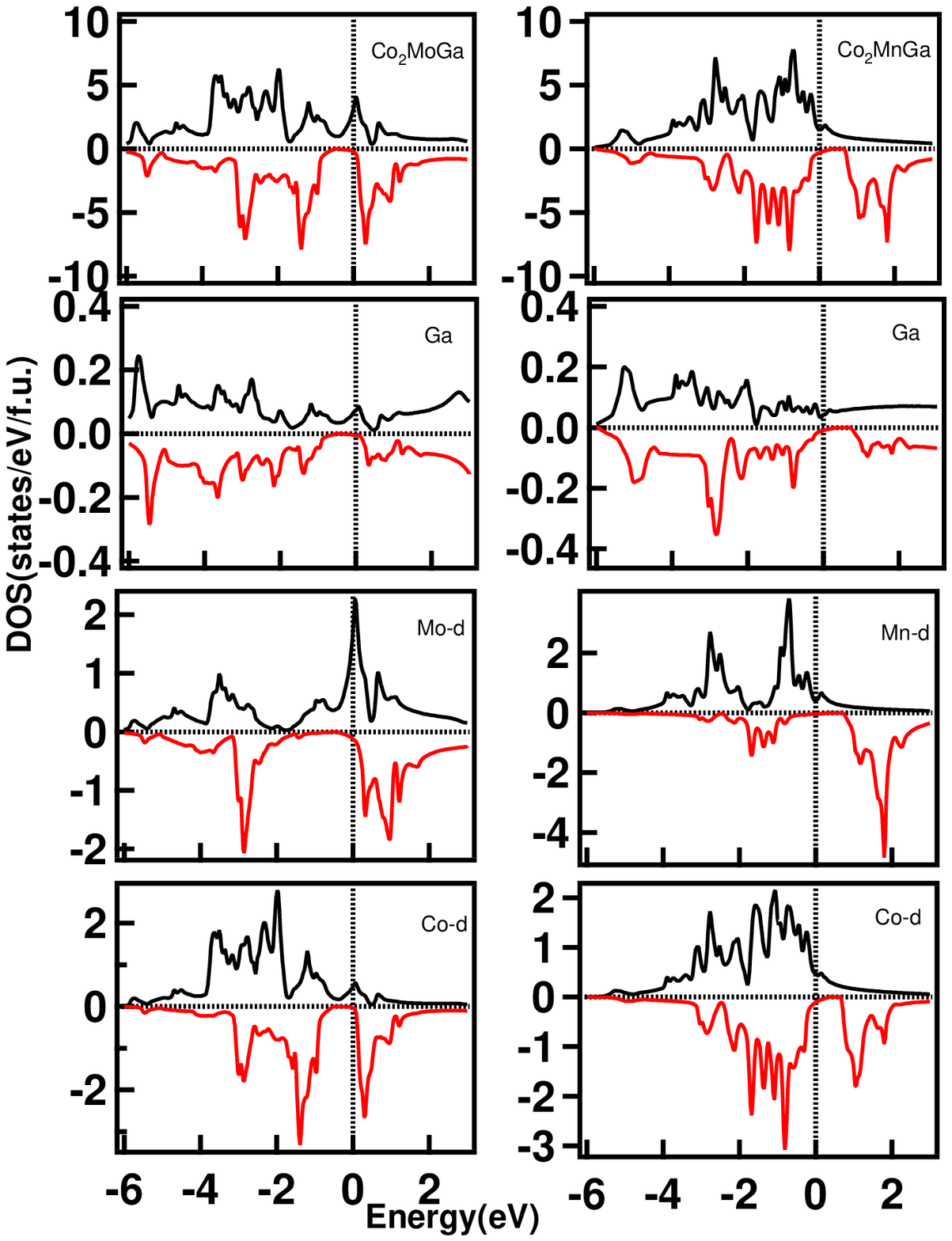}
\caption{
The left 
and right set of panels depict the density of states of Co$_{2}$MoGa 
and Co$_{2}$MnGa materials, respectively. 
From top to bottom panel: first the total density of states
as a function of energy has been plotted. 
Next panel shows the partial density of states of the Co atom. 
Partial density of states of the $B$ atom and the Ga atom are 
shown in the third and the fourth panels. 
The DOS of the $C$ atom is multiplied by a factor of 10.
The Fermi level is at 0 eV.
}
\label{fig:11}
\end{figure}

\clearpage
\pagebreak

\begin{figure}
\centering
\includegraphics{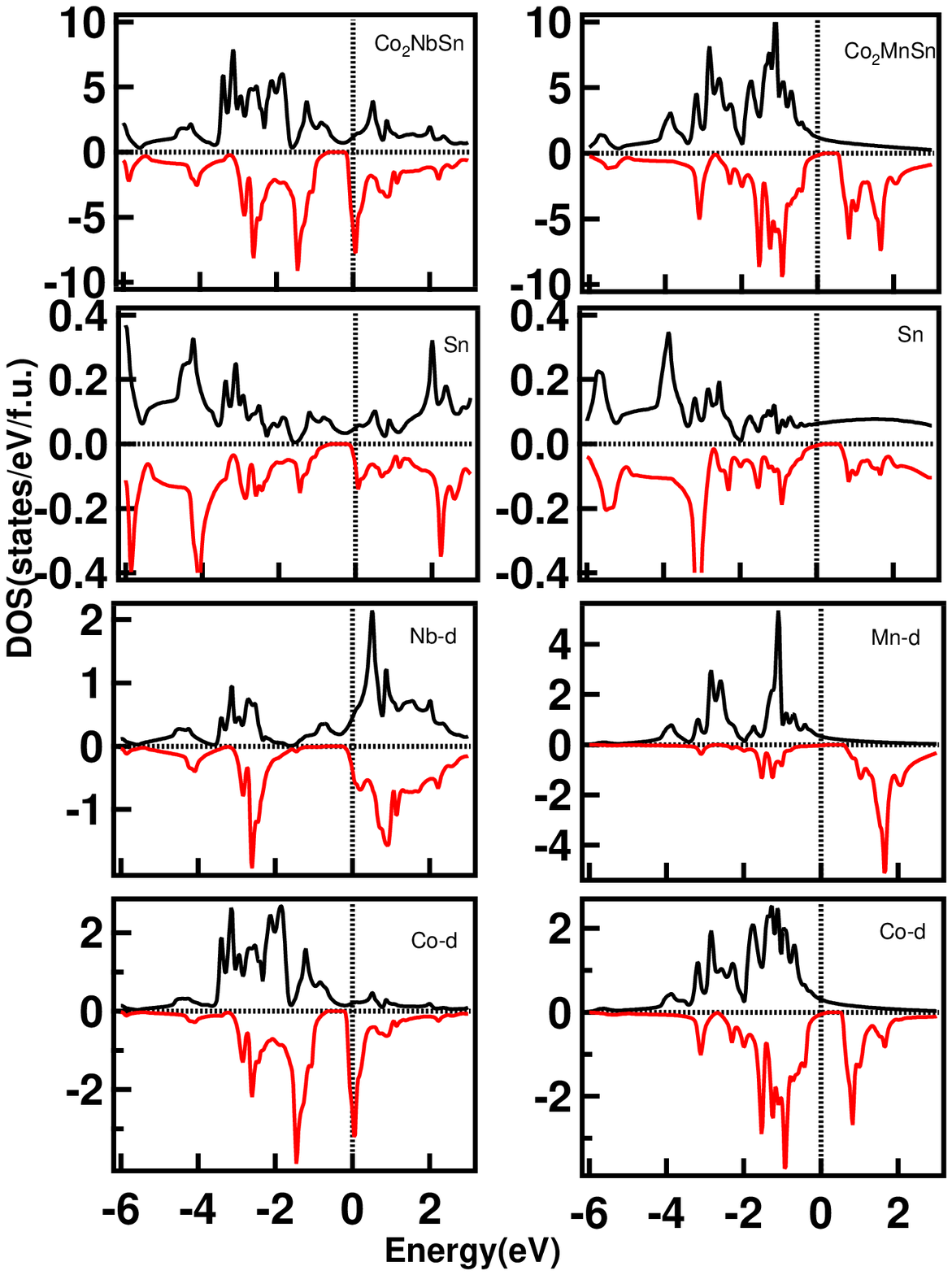}
\caption{
The left 
and right set of panels depict the density of states of Co$_{2}$NbSn 
and Co$_{2}$MnSn materials, respectively. 
From top to bottom panel: first the total density of states
as a function of energy has been plotted. 
Next panel shows the partial density of states of the Co atom. 
Partial density of states of the $B$ atom and the Sn atom are 
shown in the third and the fourth panels. 
The DOS of the $C$ atom is multiplied by a factor of 10.
The Fermi level is at 0 eV.
}
\label{fig:12}
\end{figure}

\clearpage
\pagebreak

\begin{figure}
\centering
\includegraphics{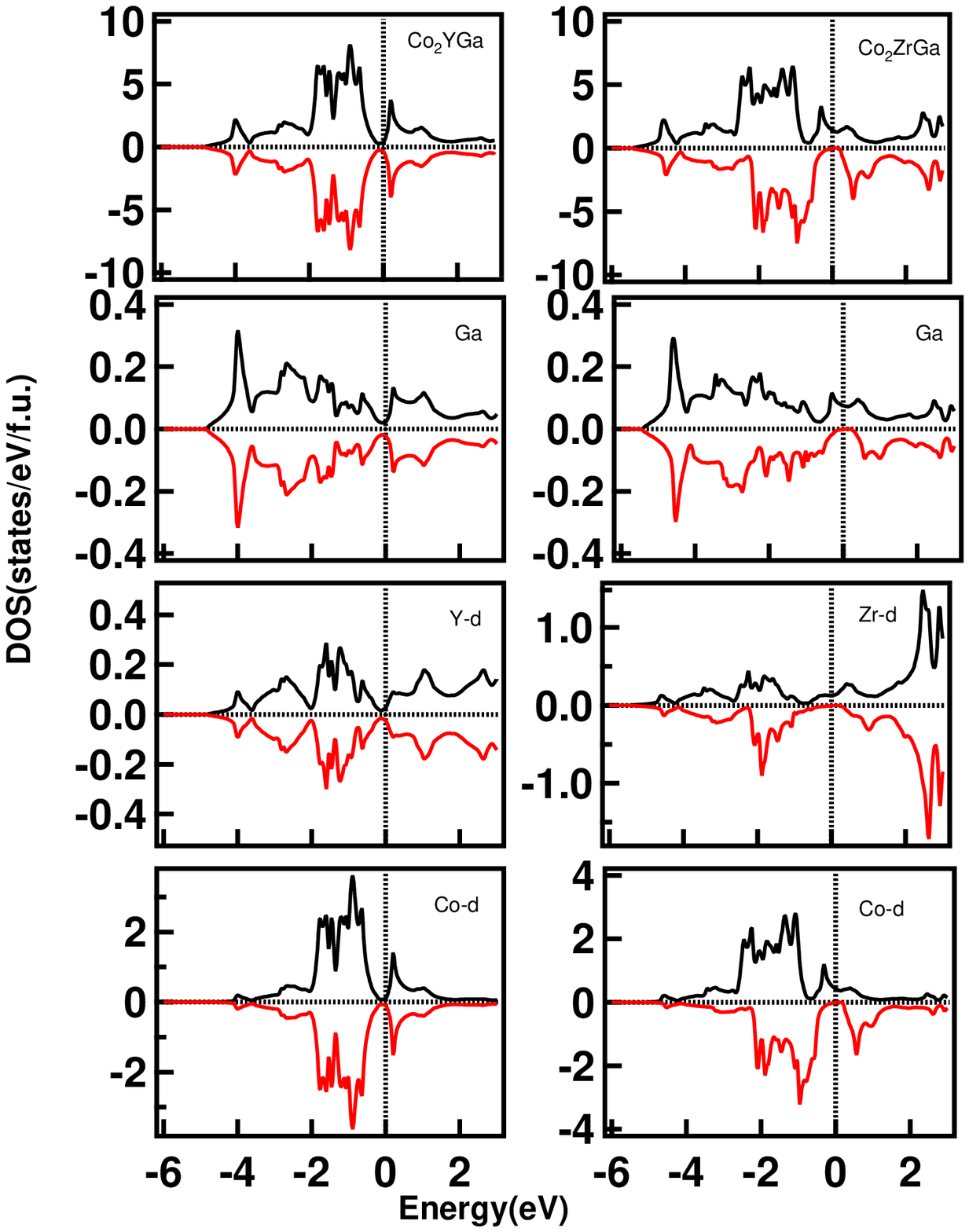}
\caption{
The left 
and right set of panels depict the density of states of Co$_{2}$YGa 
and Co$_{2}$ZrGa materials, respectively. 
From top to bottom panel: first the total density of states
as a function of energy has been plotted. 
Next panel shows the partial density of states of the Co atom. 
Partial density of states of the $B$ atom and the Ga atom are 
shown in the third and the fourth panels. 
The DOS of the $C$ atom is multiplied by a factor of 10.
The Fermi level is at 0 eV.
}
\label{fig:13}
\end{figure}

\clearpage
\pagebreak

\begin{figure}
\centering
\includegraphics{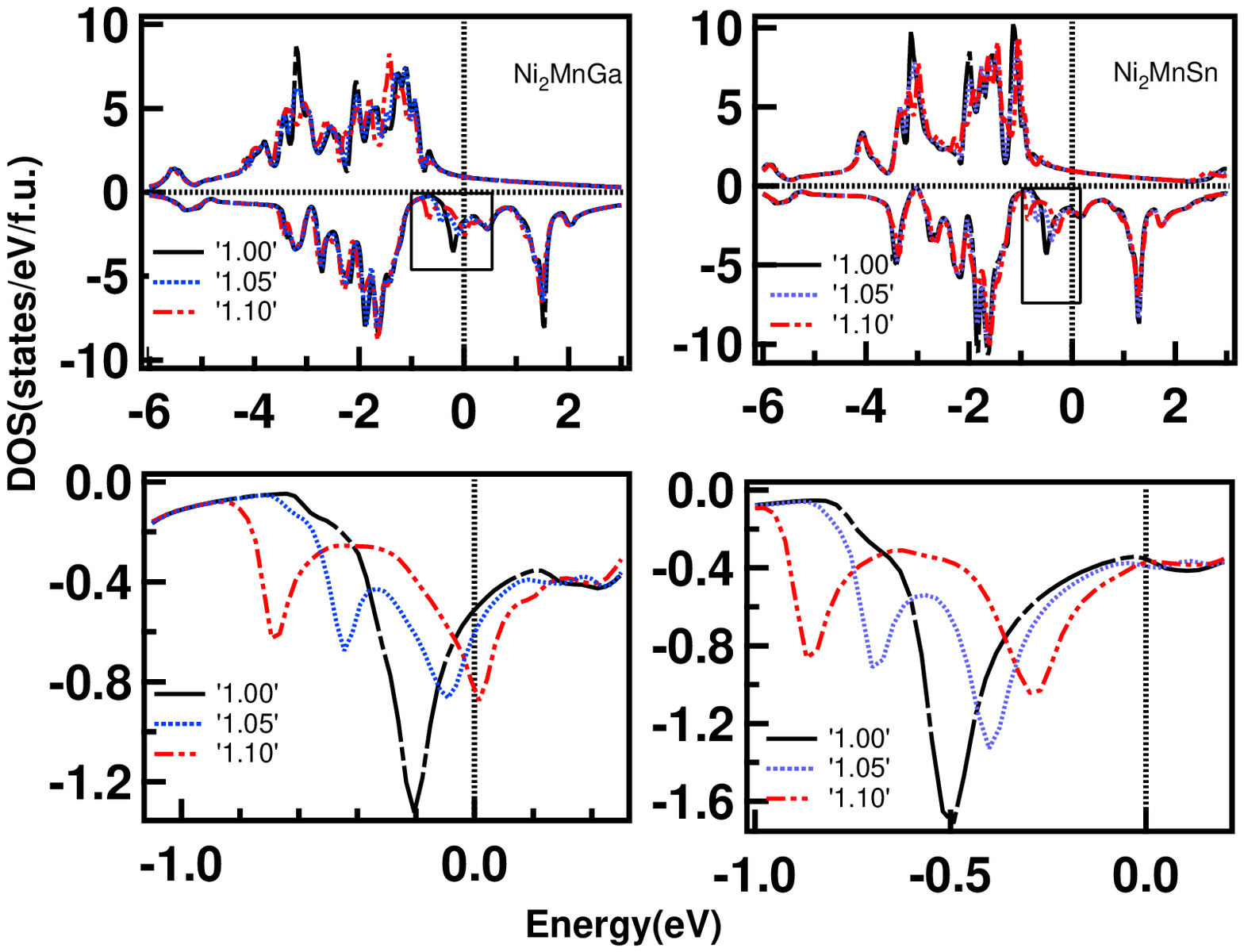}
\caption{
The density of states as a function of energy has been plotted for 
the cubic and tetragonal phases, with $c$/$a$ varying from 1 to 1.10
in steps of 0.05 for materials Ni$_{2}$MnGa and Ni$_{2}$MnSn in 
left and right panels, respectively. Panels below show the down spin 
density near the Fermi level in an expanded scale for respective
materials.
The Fermi level is at 0 eV.
}
\label{fig:14}
\end{figure}

\clearpage
\pagebreak

\begin{figure}
\centering
\includegraphics{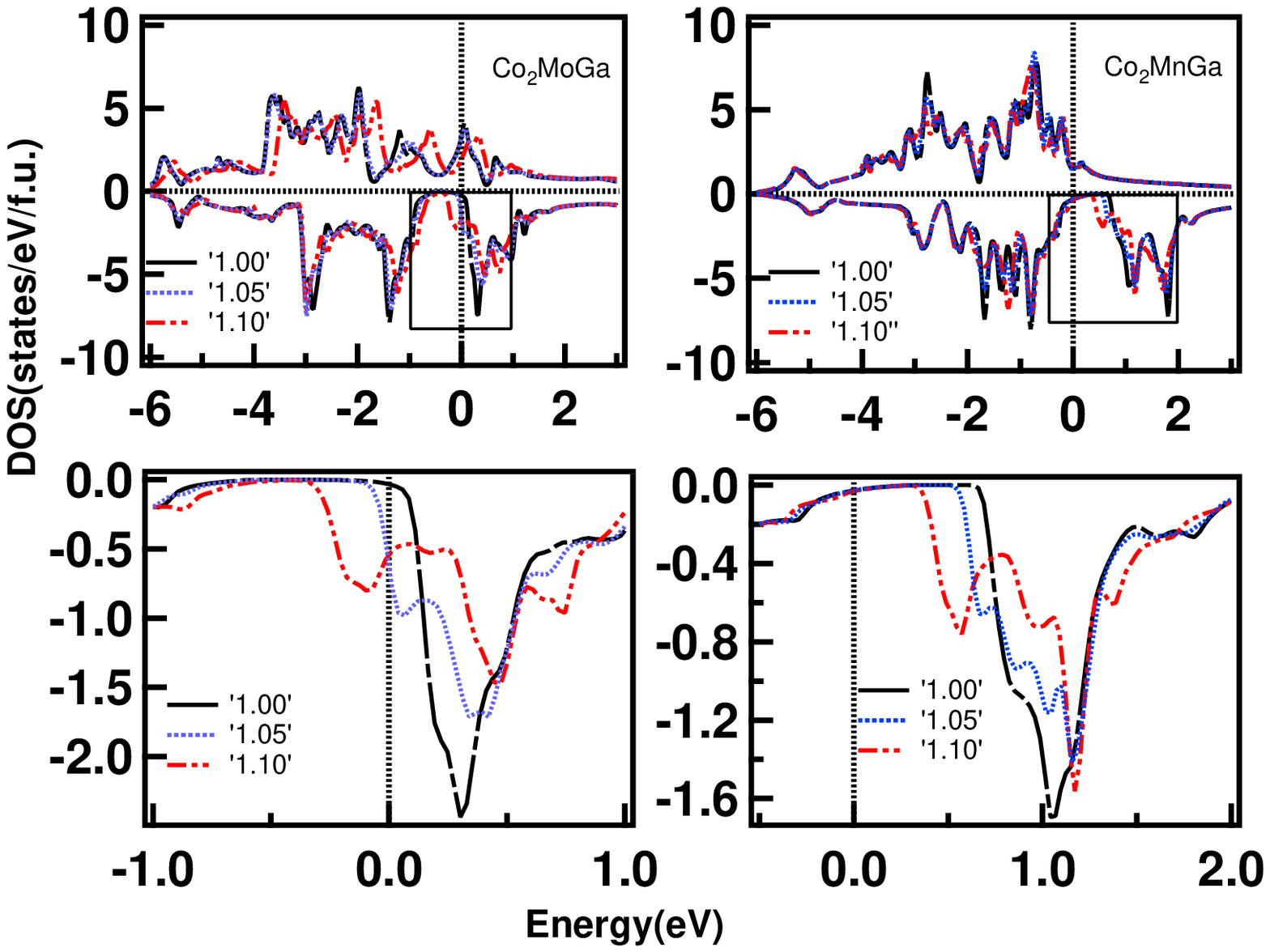}
\caption{
The density of states as a function of energy has been plotted for 
the cubic and tetragonal phases, with $c$/$a$ varying from 1 to 1.10
in steps of 0.05 for materials Co$_{2}$MoGa and Co$_{2}$MnGa in 
left and right panels, respectively. Panels below show the down spin 
density near the Fermi level in an expanded scale for respective
materials.
The Fermi level is at 0 eV.
}
\label{fig:15}
\end{figure}

\clearpage
\pagebreak

\begin{figure}
\centering
\includegraphics{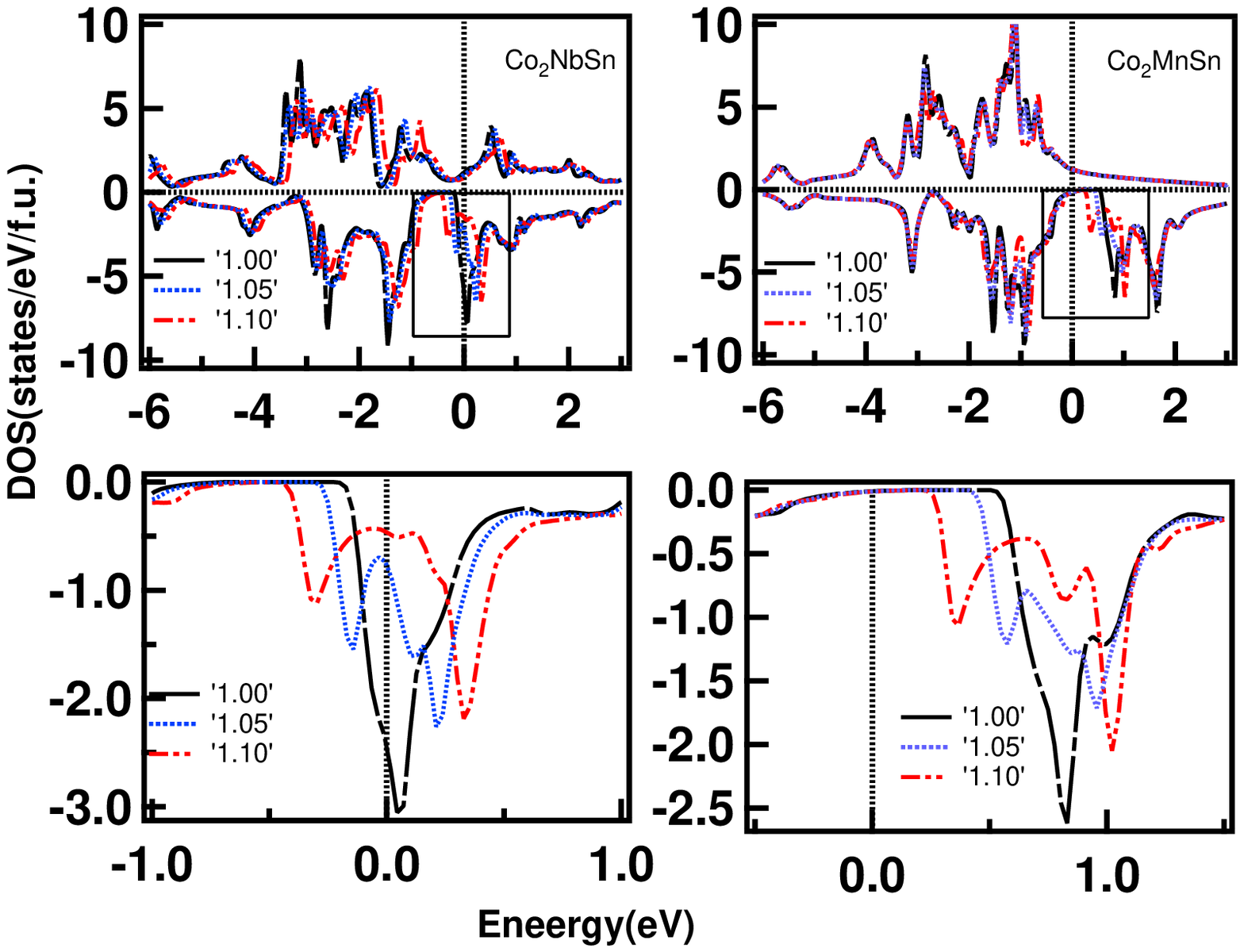}
\caption{
The density of states as a function of energy has been plotted for 
the cubic and tetragonal phases, with $c$/$a$ varying from 1 to 1.10
in steps of 0.05 for materials Co$_{2}$NbSn and Co$_{2}$MnSn in 
left and right panels, respectively. Panels below show the down spin 
density near the Fermi level in an expanded scale for respective
materials.
The Fermi level is at 0 eV.
}
\label{fig:16}
\end{figure}

\clearpage
\pagebreak

\begin{figure}
\centering
\includegraphics{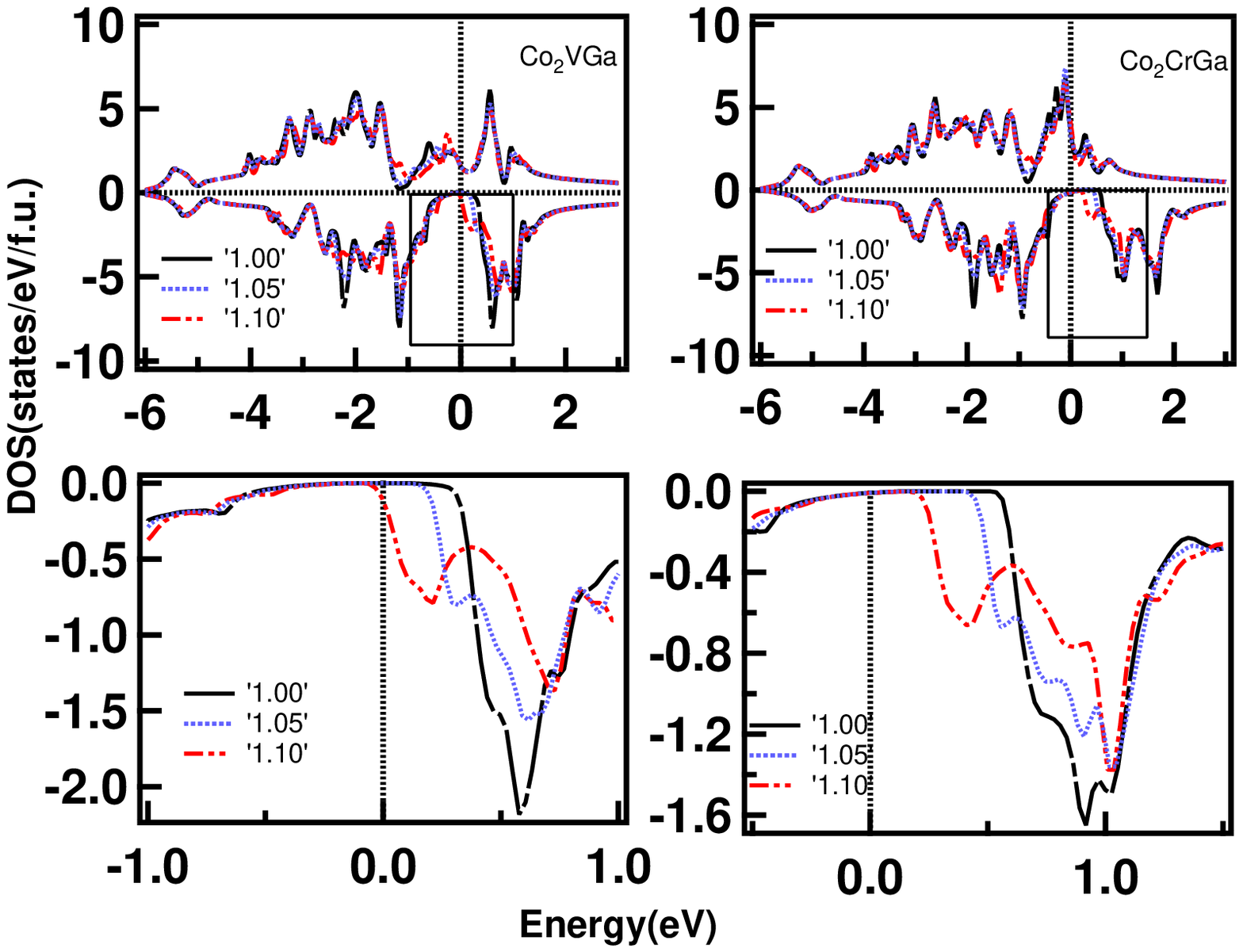}
\caption{
The density of states as a function of energy has been plotted for 
the cubic and tetragonal phases, with $c$/$a$ varying from 1 to 1.10
in steps of 0.05 for materials Co$_{2}$VGa and Co$_{2}$CrGa in 
left and right panels, respectively. Panels below show the down spin 
density near the Fermi level in an expanded scale for respective
materials.
The Fermi level is at 0 eV.
}
\label{fig:17}
\end{figure}

\end{document}